\documentclass[10pt,preprint,eqsecnum]{aastex}

\usepackage{amsmath}

\begin{document}

\title{COSMOS: A Radiation-Chemo-Hydrodynamics Code for Astrophysical Problems}

\author{Peter Anninos, P. Chris Fragile, and Stephen D. Murray}
\affil{University of California,
Lawrence Livermore National Laboratory, Livermore, CA 94550}


\begin{abstract}
We have developed a new massively-parallel radiation-hydrodynamics
code (Cosmos) for
Newtonian and relativistic astrophysical problems that also includes
radiative cooling, self-gravity,
and non-equilibrium, multi-species chemistry. Several numerical methods
are implemented for the hydrodynamics, including options for
both internal and total energy conserving schemes.
Radiation is treated using flux-limited
diffusion.  The chemistry incorporates 27 reactions,
including both collisional and radiative processes for
atomic hydrogen and helium gases, and molecular hydrogen chains.
In this paper we discuss the equations and present 
results from test problems carried out to verify the
robustness and accuracy of our code in the Newtonian regime.  An earlier
paper presented tests of the relativistic capabilities of Cosmos. 
\end{abstract}

\keywords{diffusion --- hydrodynamics --- instabilities --- 
methods: numerical --- shock waves}

\section{Introduction}
\label{sec:introduction}

This is the second in a series of papers using a new code
(Cosmos) that we have developed for a broad range of astrophysical problems, 
including scalar- and radiation-field dominated processes in the
early universe, cosmological structure formation, black-hole accretion,
neutron-star binaries, astrophysical jets, and multiphase galactic
dynamics.  Such an ambitious range of topics requires that the code be
able to accurately model a wide variety of physical processes, in both the
relativistic and Newtonian regimes.  Our first contributions therefore 
present the results of tests designed to examine the abilities of
Cosmos to compute physical processes relevant for astrophysical problems.

Cosmos is a collection of massively parallel,
multi-dimensional, multi-physics solvers utilizing the MPI paradigm
for parallel computing of
both Newtonian and general relativistic astrophysical problems.
It currently includes several different options for computational
fluid dynamics (CFD) methods, equilibrium and non-equilibrium
primordial chemistry with 27 atomic and molecular reactions,
various radiative cooling processes, 
nonequilibrium radiation flux-limited diffusion, radiation pressure,
relativistic scalar fields, Newtonian external and self-gravity, 
arbitrary spacetime curvature in the form of a generic background metric,
and viscous stress in a fully covariant formulation. 
In general, Cosmos assumes no particular symmetry in the equations,
and is therefore designed to run on structured Cartesian meshes.
However, due to the covariant formulation adopted for the
hydrodynamics equations, Cosmos can also be run on various grid
geometries (e.g., Cartesian, cylindrical, spherical) for problems
using the relativistic fluid dynamics solvers.

Numerical methods used to solve the hydrodynamics
equations include a total variation diminishing (TVD)
Godunov solver for Newtonian flows using
Roe's \citep{Roe81} approximate Riemann solver and a third order
Runge-Kutta time-marching scheme \citep{Shu88}.  
For either Newtonian or general
relativistic systems, two artificial viscosity \citep{vonNeumann50}
algorithms are available:
a non-staggered grid method in which all variables are located
at the zone-center, and a staggered grid and field centering
method similar to the Zeus code \citep{SN92} in which scalar
(vector) quantities are located at the zone (face) centers.
Two additional options for CFD methods are included for
both Newtonian and relativistic problems that are based on non-oscillatory
central difference schemes \citep{Jiang98}, also differentiated by
grid centering: staggered versus centered in time as well as space.
The relativistic hydrodynamics algorithms have been
presented in \cite{Anninos03} along with several tests of
the code, including relativistic shock tube, wall shock, and
dust accretion problems. Here we emphasize the Newtonian
and multi-physics descriptions and code tests of 
primordial chemistry, radiative cooling, and radiation diffusion
coupled together with hydrodynamics. 

We proceed in \S\ref{sec:equations} 
by describing the basic equations used by Cosmos,
and in \S\ref{sec:tests} by presenting tests of the code.
The tests are designed to examine the ability
of the code to follow shocks, blast waves and dynamical instabilities, to
advect materials, to resolve heating and cooling flows, to transfer radiation,
to simulate chemical networks, and to calculate self-gravitating gas
distributions.

\section{Basic Equations}
\label{sec:equations}

The Newtonian multi-species mass, momentum and energy
(hydrodynamic and radiation diffusion) continuity
equations are written in an Eulerian frame as
\begin{eqnarray}
 \frac{\partial \rho}{\partial t} +\nabla_i(\rho v^i)
        &=& 0,
        \label{eqn:dens} \\
 \frac{\partial \rho^{[m]}}{\partial t} +\nabla_i(\rho^{[m]} v^i)
        &=& {\sum_{i=1}^{N_{s}}}{\sum_{j=1}^{N_{s}}} 
                            {{k_{ij}(T)}{\rho^{[i]}}{\rho^{[j]}}}
            + \sum_{i=1}^{N_{s}} {{I_i(\nu)}{\rho^{[i]}}},
        \label{eqn:dens_m} \\
 \frac{\partial (\rho v^k)}{\partial t} +\nabla_i(\rho v^k v^i)
        &=& -\nabla_k (P + P_R) - \rho \nabla_k \phi , 
        \label{eqn:mom} \\
 \frac{\partial e}{\partial t} +\nabla_i(e v^i)
        &=& -P \nabla_i v^i
            +c\rho (\sigma_a E - \sigma_p a_r T^{4}) 
            + \Lambda(T,\rho^{[m]}),
        \label{eqn:en} \\
 \frac{\partial E}{\partial t} + \nabla_i(E v^i)
        &=& \nabla_i\left(\frac{c}{3\rho\sigma_r}\nabla^i E\right)
            - \frac{E}{3} \nabla_i v^i
            - c \rho\sigma_a E + c \rho\sigma_p a_r T^4 ,
       \label{eqn:enrad}
\end{eqnarray}
where $v^k$ is the fluid velocity assumed to be the same
for each of the chemical species, $e$ is the fluid energy density,
$\rho^{[m]}$ are the species densities satisfying
$\rho = \sum_m \rho^{[m]}$ for the total density, 
$\phi$ is the gravitational potential obtained from
Poisson's equation $\nabla^2\phi = 4\pi G\rho$,
$P$ is the fluid pressure, 
$P_R = E/3$ is the radiation pressure,
$a_r (= 4a/c)$ is the radiation constant,
$a$ is Stephan-Boltzmann's constant,
$c$ is the speed of light,
$E$ is the radiation energy, 
$\sigma_r$, $\sigma_a$, and $\sigma_p$ are the 
Rosseland, absorption and Planck mean opacities, 
$k_{ij}(T)$ are the rate coefficients for the 2-body reactions
which are functions of fluid temperature $T$, and
$I_j$ are frequency-integrated photoionization and
dissociation rates. The summations in (\ref{eqn:dens_m})
are over the $N_{s}$ atomic and molecular species included
in the chemistry model, up to nine:
\ion{H}{1}, \ion{H}{2}, \ion{He}{1}, \ion{He}{2}, \ion{He}{3}, 
$e^-$, H$^-$, H$_2$, H$_2^+$.
A total of 27 chemical reactions are included in the full
network, which we summarize in Table \ref{tab:chemistry} for
convenience, but refer the reader to
\cite{Abel97} and \cite{Anninos97} for more complete
descriptions of the reactions, rate coefficients,
and numerical methods. 

Assuming that reasonable models are provided for the mean opacities
of the fluid, radiation
energy is coupled to the net fluid momentum and internal energy, 
accounting for non-equilibrium heating and cooling effects in the
single temperature and gray (spectral average) approximations.
However, we have not currently coupled radiation 
in a self-consistent fashion to the chemistry 
solver and ionization states of the fluid
composition. This is a much more difficult task
due to grid resolution requirements, computational time constraints, 
multi-species interactions, multi-group transport, and non-LTE
effects expected to be important in some of our applications.
We will address these problems in future developments of Cosmos.

Three radiative cooling and heating models are implemented
in the optically thin limit
for $\Lambda(T,\rho^{[m]})$ in (\ref{eqn:en}), depending
upon whether the chemistry is known.
First, if chemistry is not solved, the cooling function is
set proportional to the square of the total number density 
with an empirical model for the ionization fraction that
approximates linearly the equilibrium result (\ref{eqn:fioneq})
described below. In particular,
\begin{equation}
\Lambda(T,\rho) = 
    \left[\sum_i \dot{e}_i(T) (f_I \rho)^2
    + \dot{e}_{M}(T) (f_M \rho)^2
    + \Lambda_{C}\right] 
    \times \left\{ \begin{array}{ll}
                   e^{(T-T_{min})/\delta T} & \mbox{if} ~ T \le T_{min} , \\
                   1 & \mbox{otherwise} ,
                   \end{array}
           \right.
\label{eqn:coolingnochem}
\end{equation}
where $\dot{e}_M(T)$ is the temperature-dependent cooling rate from metals
(including carbon, oxygen, neon, and iron lines taken from 
Bohringer and Hensler (1989),
with a metallicity dial to scale relative
to solar abundance), $\dot{e}_i(T)$ is the cooling rate from
hydrogen and helium lines,
$f_M$ is the mass fraction of metals,
$f_I$ is an estimate of the ionization fraction defined as
$\mbox{min}(1,~\mbox{max}(0,~(T_{eV} - T_c)/3) )$ with $T_c=1 eV$
to match roughly the upper and lower bounds in the equilibrium
model described below. Also,
\begin{equation}
T = \frac{m_p (\Gamma-1)}{k_B (1+f_I)} \frac{\mu e }{\rho}
\end{equation}
is the gas temperature in Kelvin, $\mu$ is the
mean molecular weight, $m_p$ is the proton mass, 
$k_B$ is Boltzmann's constant, 
and the exponential is introduced to suppress cooling
at low temperature ($T_{min}$ is typically set to $10^4~K$,
with width $\delta T \sim 500~K$).
$\Lambda_{C}$ represents Compton cooling (or heating)
due to the interaction of free electrons with the cosmic 
microwave background.

The second model applies when the chemistry is solved in
equilibrium and the ionization fraction is determined
from equating the dominant hydrogen recombination and collisional
rates \citep{Bond84,Anninos94}
\begin{equation}
\frac{f}{1-f} = 3.2\times 10^4 T_{eV}^{1.22} e^{-I_{\mathrm H}/T_{eV}} ,
\label{eqn:fioneq}
\end{equation}
where $T_{eV}$ is the temperature in electron-volts,
$I_{\mathrm H}=13.6~eV$ is the ionization energy of hydrogen.
Assuming the electrons and ions have the same temperature,
the gas pressure and energy can be written
\begin{eqnarray}
p &=& p_I + p_e = (1+f) \frac{\rho k_B T}{\mu m_p} , \\
e &=& \frac{1+f}{\Gamma-1} \frac{\rho k_B T}{\mu m_p} + 
                           \frac{\rho f k_B \widetilde{k} I_{\mathrm H}}{\mu m_p} ,
\end{eqnarray}
with equation of state
\begin{equation}
p = \left(\Gamma-1\right) 
    \left(e - \frac{\rho f k_B \widetilde{k} I_{\mathrm H}}{\mu m_p}\right) ,
\end{equation}
where $\widetilde{k} (= 11605~K/eV)$ is a numerical constant converting
between electron-volts and Kelvin.
The cooling function takes the same form as
(\ref{eqn:coolingnochem}), but in this case the ionization
fraction $f_I$ is computed iteratively from the equilibrium model $f_I=f$
of equation (\ref{eqn:fioneq}), and
the exponential cutoff is not applied since this behavior
is contained implicitly in $f$.

Equilibrium chemistry is
solved by setting the time derivatives in equations (\ref{eqn:dens_m}) to zero
(but only for the local chemistry update, transport is still accounted for
with operator splitting)
and solving the following algebraic equations,
after first making a guess for the
electron number density
\begin{eqnarray}
n_\mathrm{HI}    &=& \frac{k_2 n_\mathrm{H}}{k_1+k_2+k_{20}/n_e}, \label{eqn:eqHI} \\
n_\mathrm{HII}   &=& n_\mathrm{H} - n_\mathrm{HI}, \\
n_\mathrm{HeII}  &=& \frac{yn_\mathrm{H}}{1+k_4/(k_3+k_{21}/n_e)+k_5/k_6+k_{22}/(k_6 n_e)}, \\
n_\mathrm{HeI}   &=& \frac{k_4 n_\mathrm{HeII}}{k_3+k_{21}/n_e}, \\
n_\mathrm{HeIII} &=& n_\mathrm{HeII} \left(\frac{k_5}{k_6} + \frac{k_{22}}{k_6 n_e}\right), \\
n_{e}     &=& n_\mathrm{HII} + n_\mathrm{HeII} + 2 n_\mathrm{HeIII} , \label{eqn:eqne}
\end{eqnarray}
where $f_\mathrm{H}$ is the mass fraction of hydrogen,
$n_\mathrm{H} = f_\mathrm{H} \rho/m_p$, and $y=(1-f_\mathrm{H})/(4 f_\mathrm{H})$
parameterizes the relative helium concentration.
Equations (\ref{eqn:eqHI}) - (\ref{eqn:eqne}) are solved iteratively in the
order they are written until the electron density converges to a specified
tolerance, typically less than $10^{-10}$.
The subscripts used for the rate coefficients $k_i$ refer to a particular
chemical reaction ordered as in Table \ref{tab:chemistry}.

The third cooling model is used when chemistry is solved dynamically
with the full nonequilibrium equations (\ref{eqn:dens_m}).
In this case, the various ionization states 
and concentration densities $\rho^{[i]}$ of each element
are calculated from the chemistry equations with a stable ordered backwards
differencing scheme \citep{Anninos97} and used explicitly in the
cooling function as
\begin{equation}
\Lambda(T,\rho^{[m]}) =
    \sum_{i=1}^{N_{s}} \sum_{j=1}^{N_{s}}  \dot{e}_{ij}(T) \rho^{[i]} \rho^{[j]}
    + \sum_{i=1}^{N_{s}} J_i(\nu) \rho^{[i]}
    + \dot{e}_M(T) (f_M \rho)^2
    + \Lambda_{C} ,  \label{eqn:coolingchem}
\end{equation}
where $\dot{e}_{ij}(T)$ are
the cooling rates from 2-body interactions between species
$i$ and $j$, and $J_i$ represents frequency-integrated photoionization
and dissociation heating.
The equation of state in this case is given by
\begin{equation}
T = \frac{(\Gamma-1)}{k_B} \frac{e}{\sum_i^{N_s} n_i}.
\end{equation}
We account for a total of eight different cooling and heating mechanisms:
collisional-excitation,
collisional-ionization,
recombination,
bremsstrahlung,
metal-line cooling (dominantly carbon, oxygen, neon, and iron),
molecular-hydrogen cooling,
Compton cooling or heating,
and photoionization heating.

Finally we note that
the following conservation equations
for the chemical concentrations of
hydrogen, helium, and charge
\begin{eqnarray}
  &&\rho_\mathrm{H} + \rho_\mathrm{H^+} + \rho_\mathrm{H^-} 
                    + \rho_\mathrm{H_2^+} + \rho_\mathrm{H_2} = \rho~f_\mathrm{H}  ,
    \nonumber \\
  &&\rho_\mathrm{He} + \rho_\mathrm{He^+} + \rho_\mathrm{He^{++}} = \rho~(1-f_\mathrm{H})  ,
    \nonumber \\
  &&\rho_\mathrm{H^+} - \rho_\mathrm{H^-} + \frac{1}{2}\rho_\mathrm{H_2^+}
                      + \frac{1}{4}\rho_\mathrm{He^+} + \frac{1}{2}\rho_\mathrm{He^{++}} = m_p~n_e  ,
    \nonumber 
\end{eqnarray}
are enforced after each computational cycle update.

\section{Numerical Methods and Tests}
\label{sec:tests}

Cosmos uses
structured, block-ordered meshes for both spatial finite differencing
and finite volume discretization methods.
Depending on the hydrodynamic algorithm, state variables
are either defined all at the zone centers (for
one of the artificial viscosity schemes and for the total
energy conserving methods - TVD Godunov and non-oscillatory
central difference schemes), or on a staggered mesh for
a second artificial viscosity method
in which scalar and tensor quantities are zone-centered
and vector variables are face-centered.
Periodic, reflection, constant-in-time, user-specified, 
outgoing, and flat (vanishing first derivative) boundary
conditions are implemented.
Both the relativistic and Newtonian hydrodynamic equations 
are solved using single or multiple step time-explicit, 
operator-split methods
with second-order spatial finite differencing.
The gravitational potential and nonequilibrium radiation diffusion equations
(discretized implicitly without the transport and compressive terms)
are solved using a collection of linear matrix solvers from
the Hypre software package developed at Lawrence Livermore
National Laboratory \citep{FalgoutURL}. Hypre includes
options for different conjugate gradient and multigrid methods with
various preconditioners to optimize performance.
We use a multigrid solver for the gravity and diffusion
tests in sections \S\ref{sec:marsh}, \S\ref{sec:polytrope}, and
\S\ref{sec:polytroperad}.

Since the main emphasis in this paper is on Newtonian hydrodynamics,
the code tests presented in the following sections 
are designed to verify our code only in that regime, along with
multi-physics (e.g. radiation and  chemistry) coupling.
We refer the interested reader to \citep{Anninos03} for discussions
of the relativistic tests and for more explicit details of the numerical
algorithms.  It is anticipated that problems run in the Newtonian regime
using Cosmos shall include microphysics (chemistry, heating, and cooling).
Such problems are most appropriately solved with one of the artificial
viscosity methods, which are written in an internal energy formulation.
All of the hydrodynamics tests shown here are therefore computed using the
staggered-mesh artificial viscosity method.  We have, however, confirmed that
results from the other algorithms yield comparable accuracy for most of
the tests. The exceptions being the non-oscillatory central
difference schemes which are more diffusive, in general,
but particularly so for the Rayleigh-Taylor and spherical polytrope tests.
Accuracy comparable to the artificial viscosity approach
can be achieved for those tests if the grid
resolution is roughly doubled.

\subsection{Shock Tube}
\label{sec:stube}

We begin testing with one of the standard problems
in fluid dynamics, the shock tube or Sod problem.  This test is characterized 
initially by two different fluid states separated by a membrane.
At $t=0$ the membrane is removed and the fluid evolves in 
such a way that five distinct regions appear in the flow: an 
undisturbed region at each end, separated by a rarefaction wave, 
a contact discontinuity, and a shock wave.  This problem provides a good test 
of the shock-capturing properties of the code since
it has an exact solution \citep{Sod78} against which numerical 
results can be compared.

\begin{figure}
\plotone{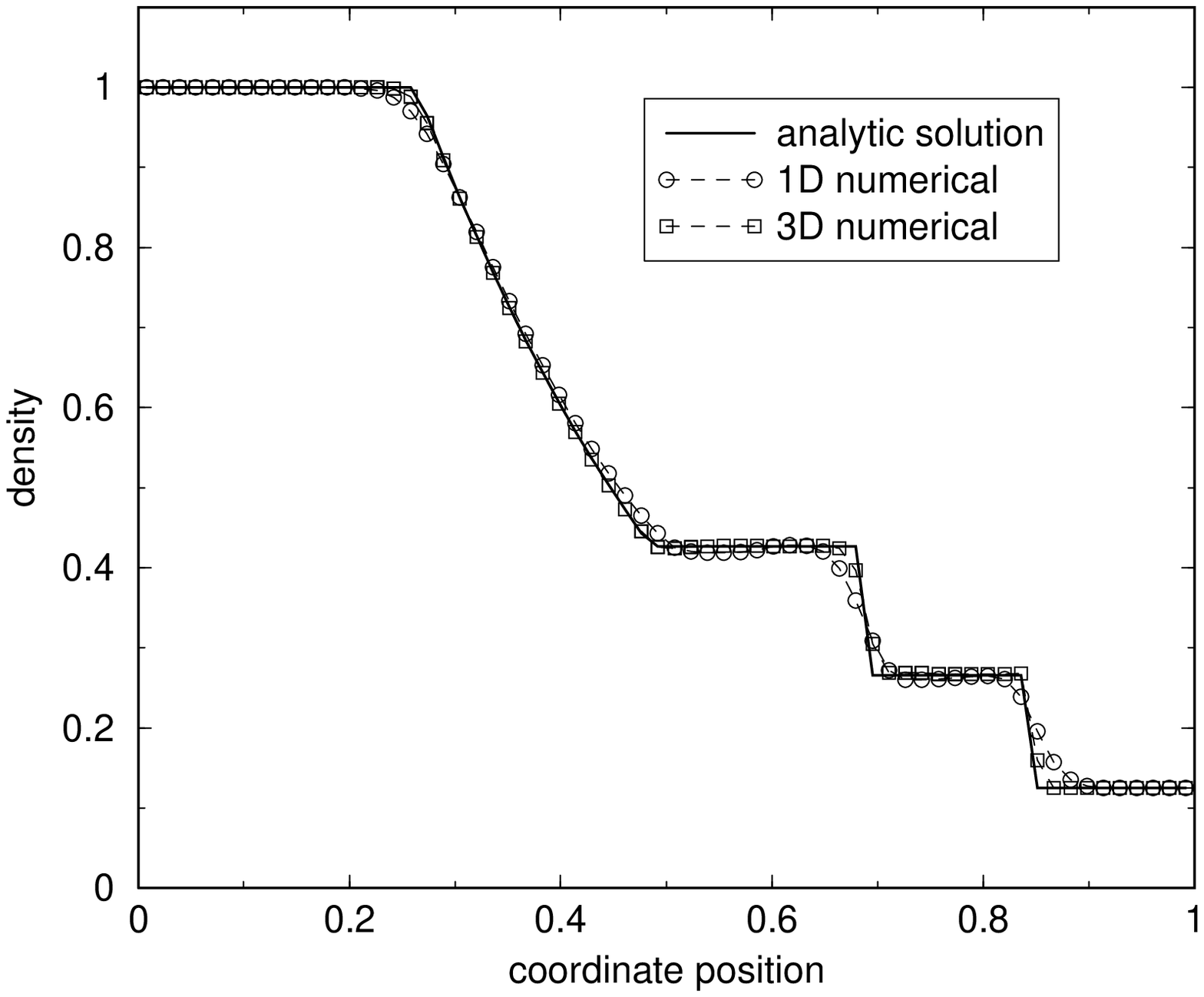}
\caption{Density profiles for the 64-zone 1D and $64^3$-zone 3D Sod tests
covering a unit grid at time $t=0.2$.
\label{fig:Sod}}
\end{figure}

The initial state is specified by $\rho_L = 1$, $P_L = 1$, and 
$V_L = 0$ to the left of the membrane, and $\rho_R = 0.125$, 
$P_R = 0.1$, and $V_R = 0$ to the right.  The fluid is 
assumed to be an ideal gas with $\Gamma = 1.4$, and the 
integration domain extends over a unit grid
from $x=0$ to $x=1$, with the membrane located at $x=0.5$.
The results presented here were run
using scalar artificial viscosity with a quadratic
viscosity coefficient $k_{q2}=2.0$, linear viscosity coefficient
$k_{q1}=0.3$, and Courant factor $k_c=0.6$, in the notation
of \cite{Anninos03}.
Figure \ref{fig:Sod} shows 
spatial profiles of the results at time $t=0.2$
for the 64-zone 1D case and the $64^3$-zone 3D case (along the main diagonal).
Table \ref{tab:Sod} summarizes the errors in $\rho$, $P$, and $V$ 
for different grid resolutions
using the $L$-1 norm (i.e.,
$\Vert E(a) \Vert_1 = \sum_{i,j,k} \Delta x_i \Delta y_j \Delta z_k
\vert a_{i,j,k}^n - A_{i,j,k}^n \vert$, where
$a_{i,j,k}^n$ and $A_{i,j,k}^n$ are the numerical and exact 
solutions, respectively, and $j=k=\Delta y_j = \Delta z_k =1$ for
1D problems).  The convergence rates for these tests are just under
first order, as expected for shock-capturing methods.  The slightly
higher $L$-1 norm errors in the 3D case are due to the fact that
the error calculation is computed globally across the whole mesh,
and so suffers from reflection effects at the grid boundaries.

\subsection{Sedov Blast Wave}
\label{sec:sedov}

The next problem we consider is the Sedov blast wave in which energy is
released at $t=0$ in the form of an explosion into an initially
undisturbed, uniform gas.  In 3D, this results in a spherical shock wave
(or blast wave) expanding from the explosion, such that $r_S \propto t^{2/5}$,
where $r_S$ is the radius of the shock and $t$ is the elapsed time 
\citep{Sedov59}.  This problem
encompasses a number of useful tests for our code, as it determines how well the
code can follow a spherical shock wave as well as testing energy
conservation.

\begin{figure}
\plotone{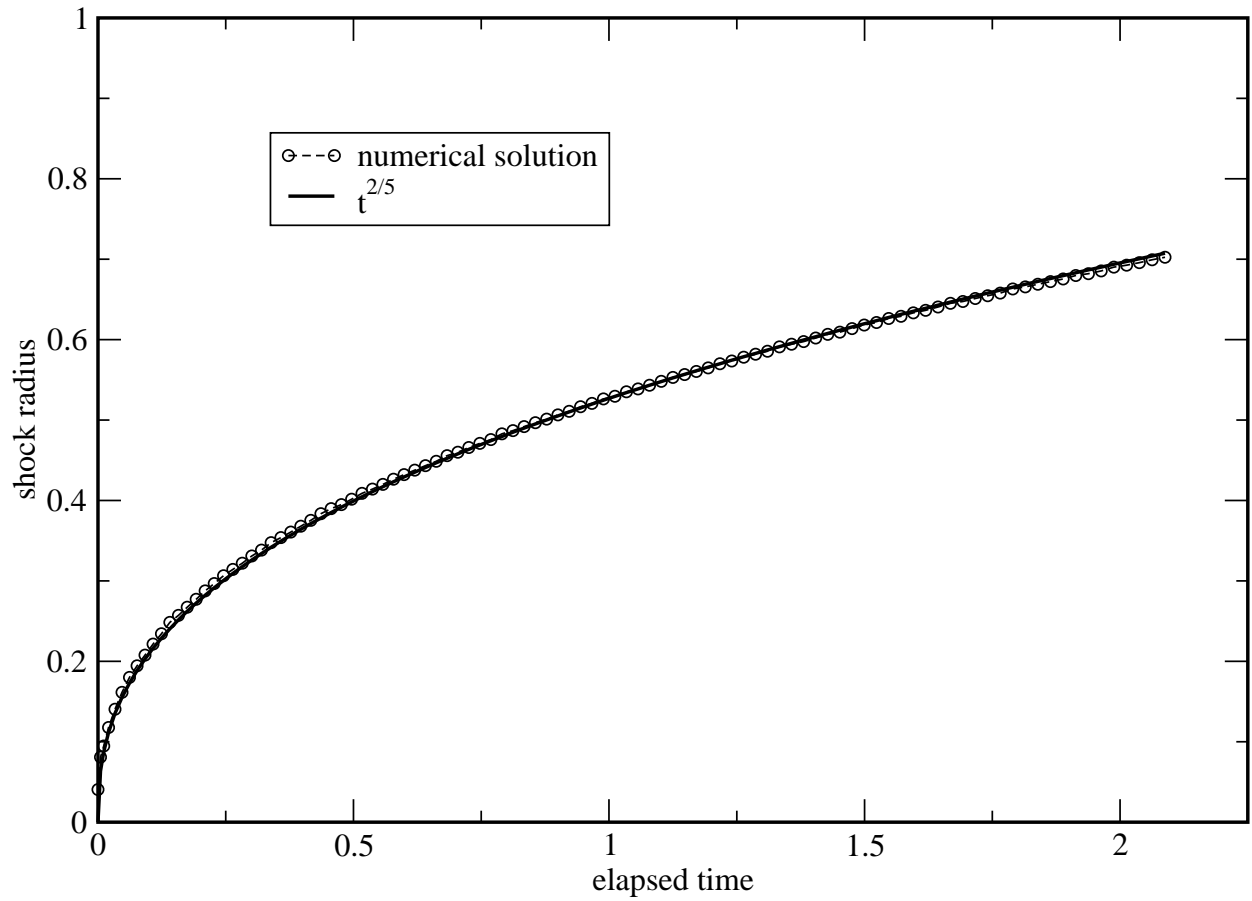}
\caption{Shock radius as a function of time for a $64^3$-zone 3D Sedov test
covering a unit grid. 
\label{fig:Sedov1}}
\end{figure}

\begin{figure}
\plotone{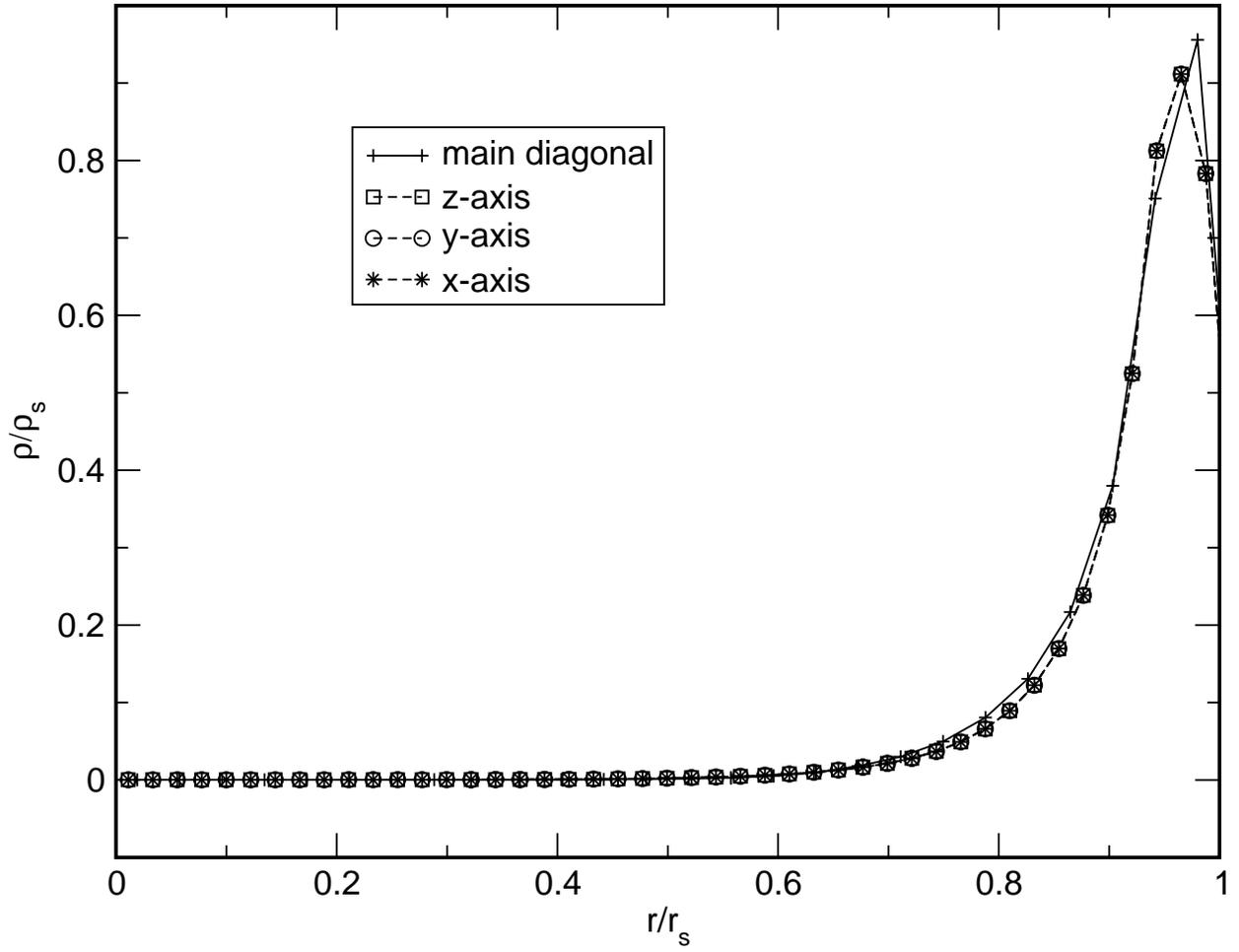}
\caption{Density distribution behind the shock wave for a $64^3$-zone 
3D Sedov test covering a unit grid at time $t=2.1$.
\label{fig:Sedov2}}
\end{figure}

The initial state is specified by $\rho_0 = 1$, $P_0 = 3.33 \times 10^{-11}$, 
and $V_0 = 0$.  The fluid is 
assumed to be an ideal gas with $\Gamma = 4/3$, and the 
integration domain extends over a unit cube.
The blast wave is initiated by significantly increasing the energy density 
(relative to the background) in an approximately
spherical region with a half-width at half-maximum of two
zones and maximum cutoff radius of five zones 
using a Gaussian profile.  The initial energy density 
contrast between the peak of the Gaussian and the
background is $6.6 \times 10^{11}$.  The results presented here were run 
using a tensor artificial viscosity with a quadratic
viscosity coefficient $k_{q2}=1.0$, linear viscosity coefficient
$k_{q1}=0.3$, and Courant factor $k_c=0.4$. 
Figure \ref{fig:Sedov1} shows the shock radius as a function of time fit
with a curve of the form $t^{2/5}$.
Figure \ref{fig:Sedov2} shows 
spatial profiles of the density along the $x$, $y$, and $z$ axes, as well
as the diagonal at time $t=2.1$ for a $64^3$-zone octant, demonstrating
that the blast wave maintains its self-similar solution along
all three major axes and the diagonal.
The energy loss in this problem was about 19\%, although much of this was
at the beginning of the simulation, and the total energy approaches a
constant value after a time $t=0.28$.

\subsection{Rayleigh-Taylor Instability}
\label{sec:RT}

The growth of a classical Rayleigh-Taylor instability has been modeled in
two-dimensions.  These models serve as tests of the ability of the code to
follow the growth of a classical instability in both the linear and non-linear
phases, and of its ability to cleanly advect material across the grid during
the non-linear growth.  

The system which we model has physical dimensions that run 
from 0 to 0.1 in $x$, and from -0.35 to 0.15 in $z$.  The resolution
is 128 zones in $x$ and 1280 in $z$.  The constant gravitational field,
$g=1$, points toward negative $z$.  At $z=0$, the density and isothermal
sound speed of the dense (upper) fluid are $\rho_u=1$ and $(c_s)_h^2=2.4$, 
while those of the light (lower) fluid are $\rho_l=0.1$ and $(c_s)_l^2=24$,
such that pressure is continuous across the fluid interface.  Away
from the interface, the individual fluids are isothermal, and their densities
vary so as to maintain hydrostatic equilibrium in the gravitational field, 
i.e.
\begin{equation}
\rho(z)=\rho(0)\exp{\left[-\phi(z)/c_s^2\right]},
\end{equation}
where $\phi(z)$ is the gravitational potential, and we take $\phi(0)=0$.
To approximate incompressibility, the fluids
are taken to be ideal gases, with large adiabatic index, $\Gamma=10$.

The initial distribution of the fluids, in which quantities depend only upon
$z$, are perturbed by introducing a vertical shift of the form
\begin{equation}
\delta z(x)=A\cos\left(2\pi x/\lambda\right)~,
\end{equation}
where $\lambda$ is the wavelength of the perturbation, and $A$ is its
amplitude.  We have examined models having two different values of
$A/\lambda$.  In the first, we choose $\lambda=0.1$, and $A=0.01$, such that
the perturbation is linear, though not strongly so.  Resolution requirements
and computational time constraints prevent us from modeling a system having a
substantially smaller value of $A$.  In the second model, $A=\lambda=0.1$,
such that this model examines the growth of initially non-linear 
perturbations.  In introducing the perturbations, density and pressure
are not altered, and so the fluids remain in pressure equilibrium with each
other (thus preventing the initial growth of sound waves), but are slightly
out of equilibrium with the gravitational field.  Reflective boundary
conditions are applied at the top and bottom of the grid, while periodic
boundaries are used at the left and right.
Artificial viscosity is not used in these calculations.

\begin{figure}
\plotone{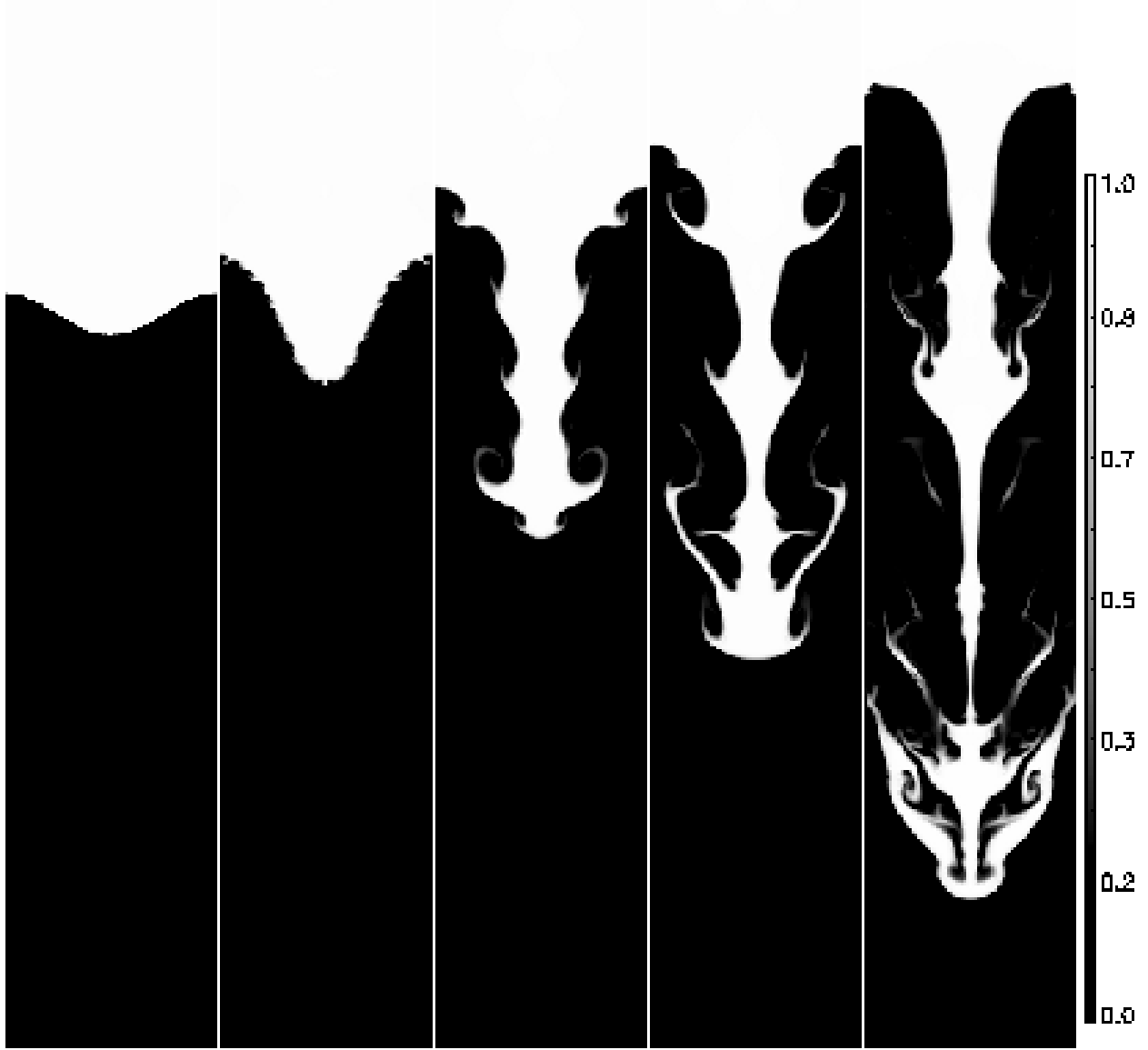}
\caption{Evolution of the Rayleigh-Taylor instability.  Shown are tracer
distributions of the dense upper fluid at times $t=$ 0.0, 0.2, 0.5, 0.7, and 1.1.}
\label{fig:RT1}
\end{figure}

Figure \ref{fig:RT1} shows snapshots of the evolution of the model having 
$A/\lambda=0.1$ at the times $t = 0, 0.2, 0.5, 0.7,$ and $1.1$.  Shown, in
grayscale, is the
tracer material initially placed in the dense upper fluid.
The initial perturbation can be seen in the first panel.
As can be seen in the subsequent panels, the discretization of the grid
creates short-wavelength structure superimposed upon the long-wavelength
perturbation.  The shortest wavelengths saturate 
quickly, but longer wavelengths persist, and can be seen superimposed upon
the classical single-mode rollup at $t=0.5$.  Although the small scale
vorticity spreads the tracer material widely across the grid, little 
diffusiveness is apparent.

\begin{figure}
\plotone{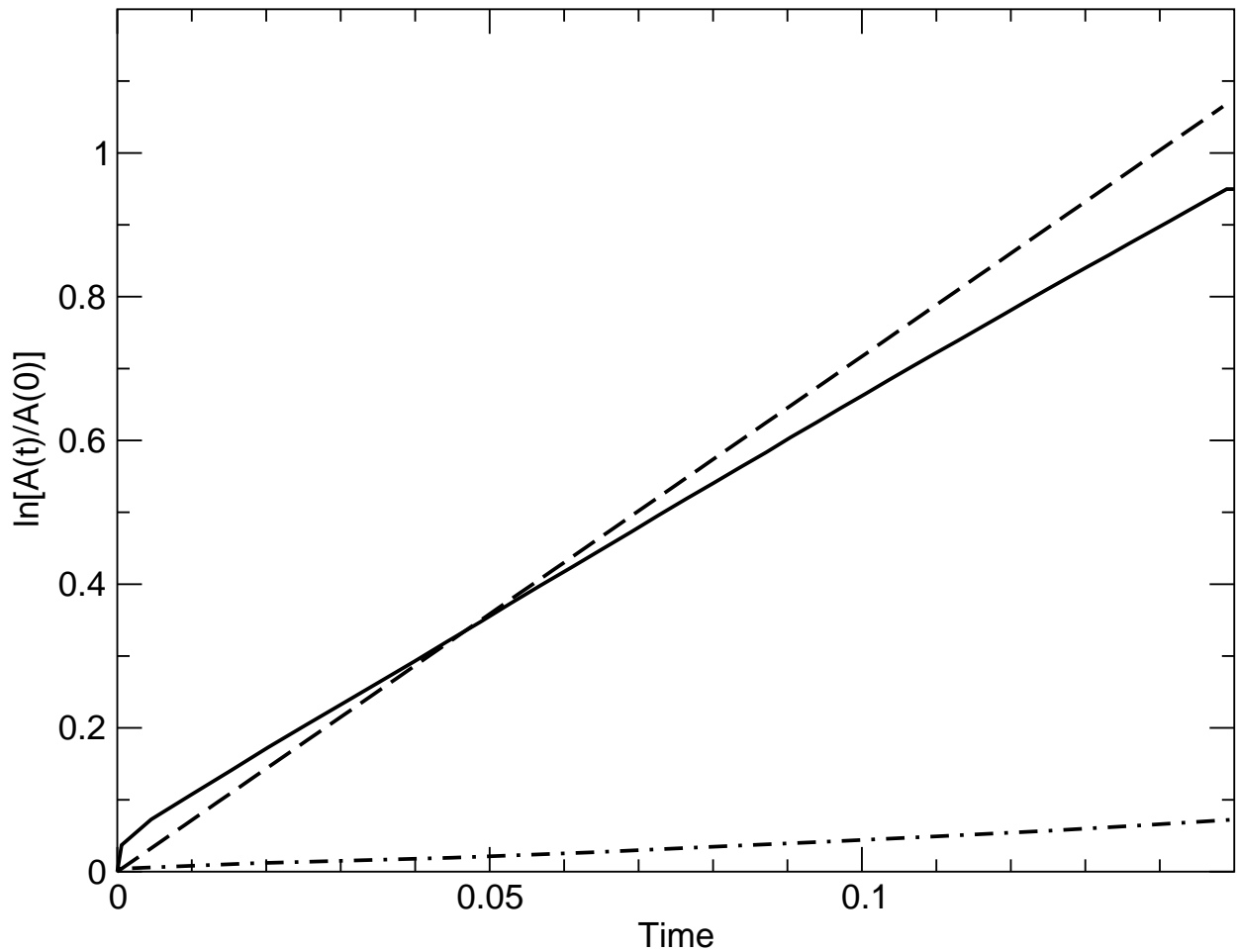}
\caption{Early growth rate for the Rayleigh-Taylor instability,
showing the evolution of the amplitude normalized to its initial value
versus time.  Results are shown for the numerical simulations having initial 
$A/\lambda=0.1$ (linear regime; solid curve) and $A/\lambda=1$ 
(non-linear regime; dash-dot curve).  The
dashed curve shows the theoretical prediction for incompressible fluids in
the linear regime.}
\label{fig:RT2}
\end{figure}

The early growth of the longest wavelength modes of both non-linear and
linear initial perturbations are shown in Figure~\ref{fig:RT2}.
Also shown is the prediction of linear theory for
incompressible, constant density fluids, for which the instability is
predicted to grow as $e^{\omega t}$, with
\begin{equation}
\omega^2=\frac{2\pi g}{\lambda} \left( \frac{\rho_u-\rho_l}{\rho_u+\rho_l}
\right) ,
\end{equation}
\citep{Chandra61}.
The overall agreement between theory and the model with the linear initial
perturbation is good.  The computed growth rate is found to be
14\% slower than the analytic value.  The slow growth rate may be due to the
relatively large
initial amplitude of the imposed perturbation, a surmise supported by the
curvature seen in the growth rate of the numerical results.  Both curves
stand in contrast to that of the initially non-linear perturbation, which has
a much smaller growth rate, as predicted by theory (see below).

\begin{figure}
\plotone{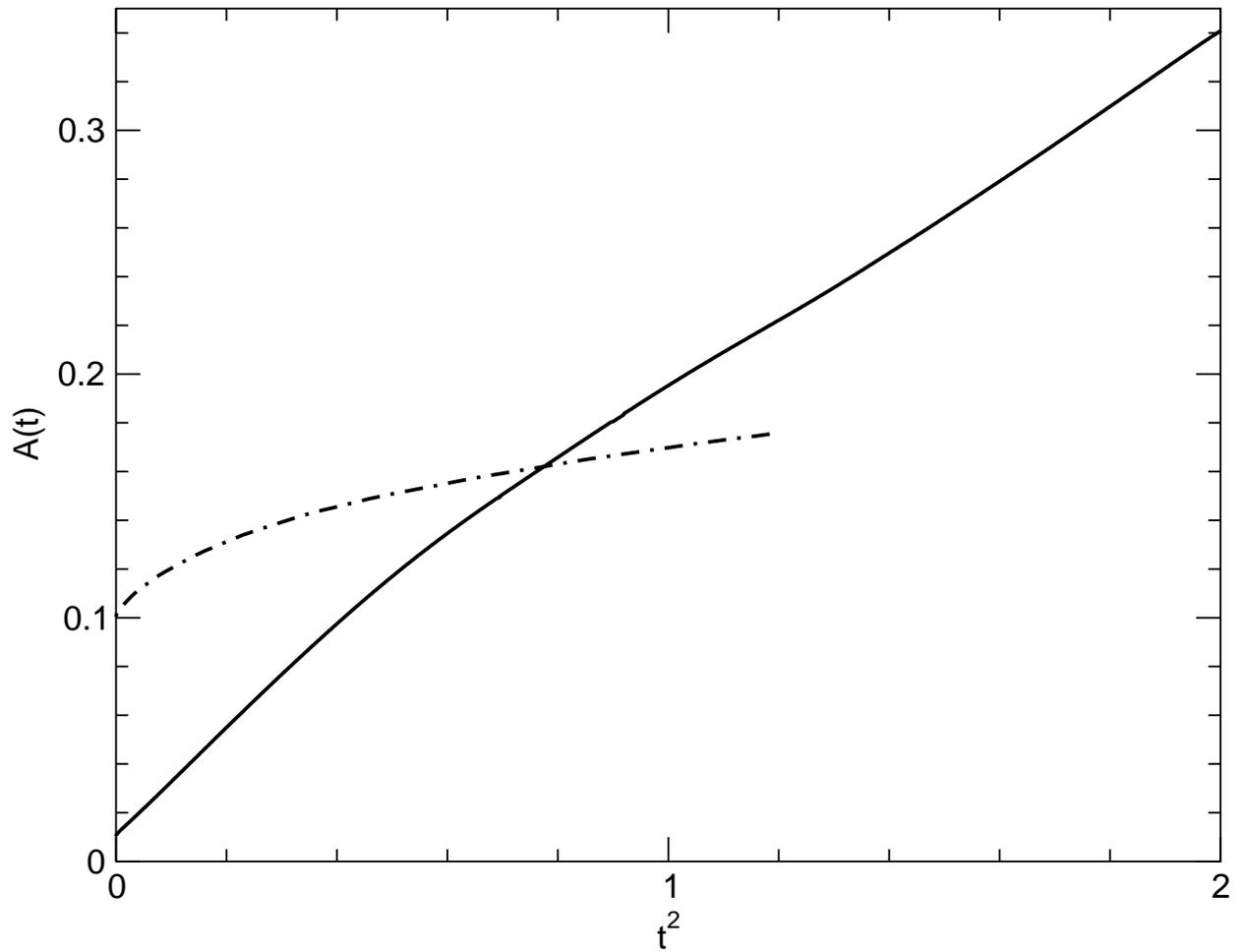}
\caption{Growth rate of the Rayleigh-Taylor instability at late times.
Shown is the amplitude of the dense fluid relative to the unperturbed
interface location for both the initially linear perturbation (solid curve),
and the initially non-linear perturbation (dot-dash curve). 
The amplitude is plotted as a function of $t^2$, to
demonstrate the asymptotic $A(t)\propto t^2$ behavior.
}
\label{fig:RT3}
\end{figure}

In the non-linear regime, the amplitude of the Rayleigh-Taylor instability is
known to behave asymptotically as 
\begin{equation}
A(t) = \alpha_s g \frac{\rho_u - \rho_l}{\rho_u + \rho_l} t^2~,
\end{equation}
\citep[e.g.][]{Youngs94,Glimm01}.  The late-time behavior of the models is
shown in Figure~\ref{fig:RT3}.  In the figure are shown the penetration
amplitude of the dense fluid as a function of $t^2$ for both the initially
linear and initially non-linear simulations.  As can be seen in the figure, 
both simulations display the expected late-time behavior, with 
$\alpha_s \approx 0.17$ for the initially linear model, and
$\alpha_s \approx 0.04$ for the initially non-linear model, where $\alpha_s$
is computed for the initial Atwood number of the fluids.  The differences
between the two models are most likely due to compressibility effects, 
which become more apparent at late times.

\subsection{Radiative Shock Waves}
\label{sec:rad}

\begin{figure}
\plotone{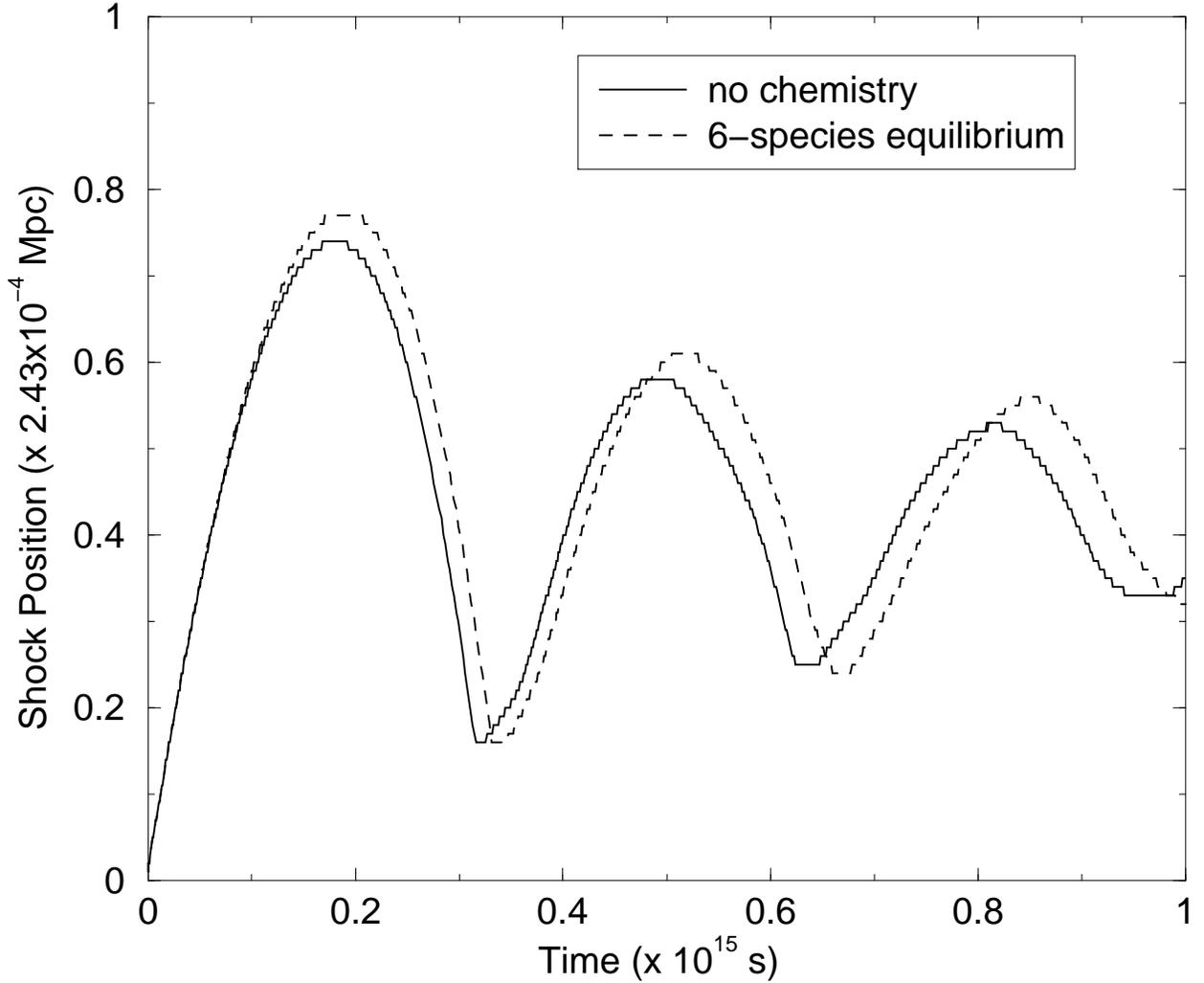}
\caption{Shock front position as a function of time in two radiative shock
simulations: one without chemistry in which the cooling function
is $\propto \rho^2 T^{1/2}$; and one where the kinetics
equations for a 6-species chemistry model
with \ion{H}{1}, \ion{H}{2}, \ion{He}{1}, \ion{He}{2}, \ion{He}{3}, and $e^-$
is solved in equilibrium (though we note that the nonequilibrium
equations yield nearly identical results as the equilibrium case).
The fundamental frequency as defined
in the text is $\omega_I = $ 0.319 and 0.315
for the no-chemistry and 6-species cases respectively. Both
compare nicely with the perturbation result of 0.31
derived by \citep{ChevIm82}.
\label{fig:radshock}}
\end{figure}

The initial data for this test is characteristic of pre-shock flows
expected from large-scale (galactic-type) structures
\begin{eqnarray}
\rho &=& 4.72 \times 10^{-25} {\rm g~cm}^{-3}~, \nonumber \\
e &=& 1.0 \times 10^{-30} {\rm g~cm}^{-1}{\rm ~s}^{-2}~, \nonumber \\
v^x &=& -u_{\rm in} = -1.7 \times 10^7 {\rm cm~s}^{-1}~, \nonumber
\end{eqnarray}
corresponding to a uniform flow of gas along the $-x$ direction.  Reflection
boundary conditions are imposed at $x=0$ and we use 100 zones to resolve
a spatial extent of $L=2.43 \times 10^{-4}$ Mpc.  
A shock wave forms at the wall at $x=0$ and propagates to the right at
velocity $v_s \sim u_{\rm in}/3$. As the heated gas cools
radiatively, the shock begins
to lose pressure support and slows down.  Eventually the shock collapses
and re-establishes a new pressure equilibrium closer to the wall.  As
gas continues to accrete, the shock front moves
outwards again to repeat the cycle of oscillations as shown
in Figure \ref{fig:radshock}, where the shock position, $x_s$, is plotted as a function
of time in units where the grid length is set to unity
and the unit of time is $10^{15}$ seconds. 

\begin{figure}
\plotone{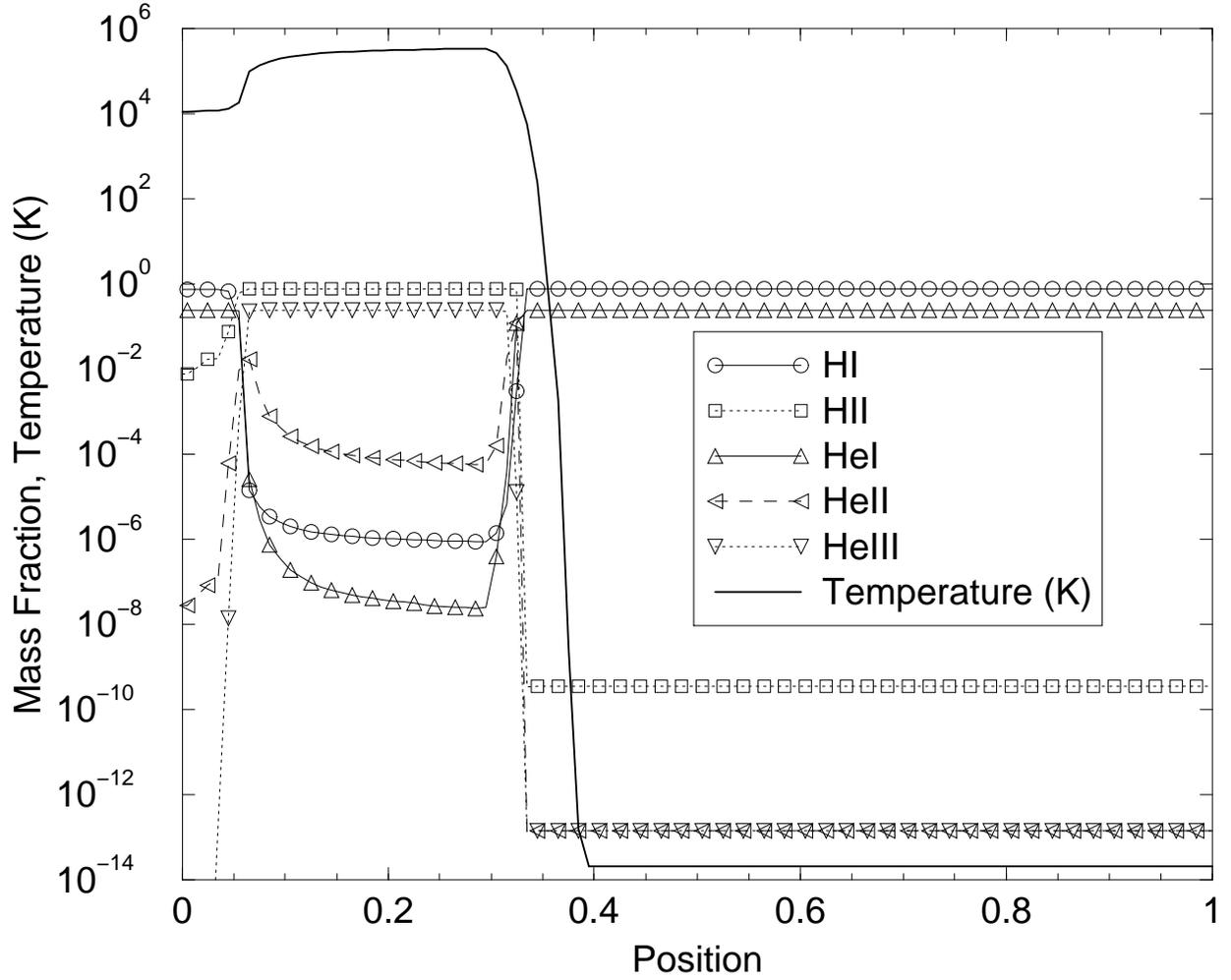}
\caption{Mass fraction distribution ($\rho_i/\rho$) of the species
\ion{H}{1}, \ion{H}{2}, \ion{He}{1}, \ion{He}{2}, and \ion{He}{3} at the
final time $t=1$ in the radiative shock test. The results shown are
from the 6-species nonequilibrium model, and agree nicely
with the equilibrium model in the hot post-shocked
phase where collisional ionization and recombination 
balance is a good approximation.
\label{fig:species}}
\end{figure}

Figure \ref{fig:radshock}
shows results from two calculations: one with no chemistry in which
$\Lambda(T,\rho)\propto \rho^2 T^{1/2}$, and a second 6-species equilibrium
chemistry model that approximates the number densities
of the dominant coolants $n_\mathrm{HI},~n_\mathrm{HII},~n_\mathrm{HeI},~n_\mathrm{HeII},~n_\mathrm{HeIII},~n_e$
from equilibrium assumptions.
We have also run a third calculation to test
the 6-species nonequilibrium chemistry model which explicitly solves the
non-linear differential kinetics equations. The results in this case are 
nearly identical to the equilibrium calculations so we do not
include them in Figure \ref{fig:radshock}.
We do, however, show the mass fraction distribution of each of the chemical
species at the final time of the simulation for the nonequilibrium
case in Figure \ref{fig:species}. Mass fractions in the hot
phase, where collisional ionization and recombination effects are
expected to be in equilibrium, agree with those computed from the equilibrium model.
For all these calculations
we assume a perfect fluid model for the gas with adiabatic index $\Gamma=5/3$
and a cooling function dominated by bremsstrahlung effects.

The accuracy of these calculations is evaluated by comparing the
fundamental frequency of oscillations to the perturbation results of
\cite{ChevIm82} who defined the normalized
frequency as $\omega_I = (2\pi/P)(\overline x_s/u_{in})$,
where $P$ is the period of the oscillations, $\overline x_s$ is the
average shock position, and $u_{in}$ is the inflow velocity.
We find $\omega_I = $ 0.319 and 0.315 for the no-chemistry and
6-species chemistry cases respectively. These results compare
nicely with the perturbation estimate of 0.31.

\subsection{Marshak Wave}
\label{sec:marsh}

In the following test, we consider the penetration of radiation from a hot 
source into cold material \citep{Marshak57}.  Because radiative transport is
very efficient in the problems considered here, 
significant penetration can occur on a timescale much shorter than 
the timescale for motion of the gas.  We therefore ignore hydrodynamic 
transport and consider only supersonic radiation diffusion through an
ambient gas in thermal equilibrium with the
radiation and held initially at temperature $T_0$, but subject to the
boundary condition $T=T_1>T_0$ at one end.

The energy diffusion equation
\begin{equation}
\frac{\partial E}{\partial t} = \nabla (D \nabla E) ~,
\end{equation}
where $D = c/(3\rho\sigma_r)$ is the diffusion coefficient,
and $\sigma_r = \kappa_0(\rho/
\rho_0)^\gamma(T/T_0)^{-m}$ is the Rosseland mean opacity,
can be solved approximately for this simple case to give \citep{Long86}
\begin{equation}
E(x,t) \approx E_1 \left\{ \xi_0^2 \left[ \frac{\bar{n}}
{\bar{n}+1} 
\left( 1-\xi/\xi_0 \right)^{(\bar{n}+1)/\bar{n}} \left(1- \frac{1-\xi/\xi_0}
{2\bar{n}+1} \right) \right] \right\} ^{1/\bar{n}} ~,
\end{equation}
where
\begin{eqnarray}
\xi   &=& \left(\frac{\bar{n}x^2}{2\eta^2 Dt} \right)^{1/2} ~, \\
\xi_0 &=& \left(\frac{\bar{n}x_f^2}{2\eta^2 Dt} \right)^{1/2}  
       = \left( \frac{(\bar{n}+1)(\bar{n} + 0.5)}{\bar{n}^2} \right)^{1/2} ~,
\label{eqn:Marsh_front}
\end{eqnarray}
$\bar{n} = (m+4)/4$, and $\eta$ is a numerical fitting factor of order unity.  
From (\ref{eqn:Marsh_front}) we clearly see that the radiation front
location, $x_f$, should propagate as $t^{1/2}$.

\begin{figure}
\plotone{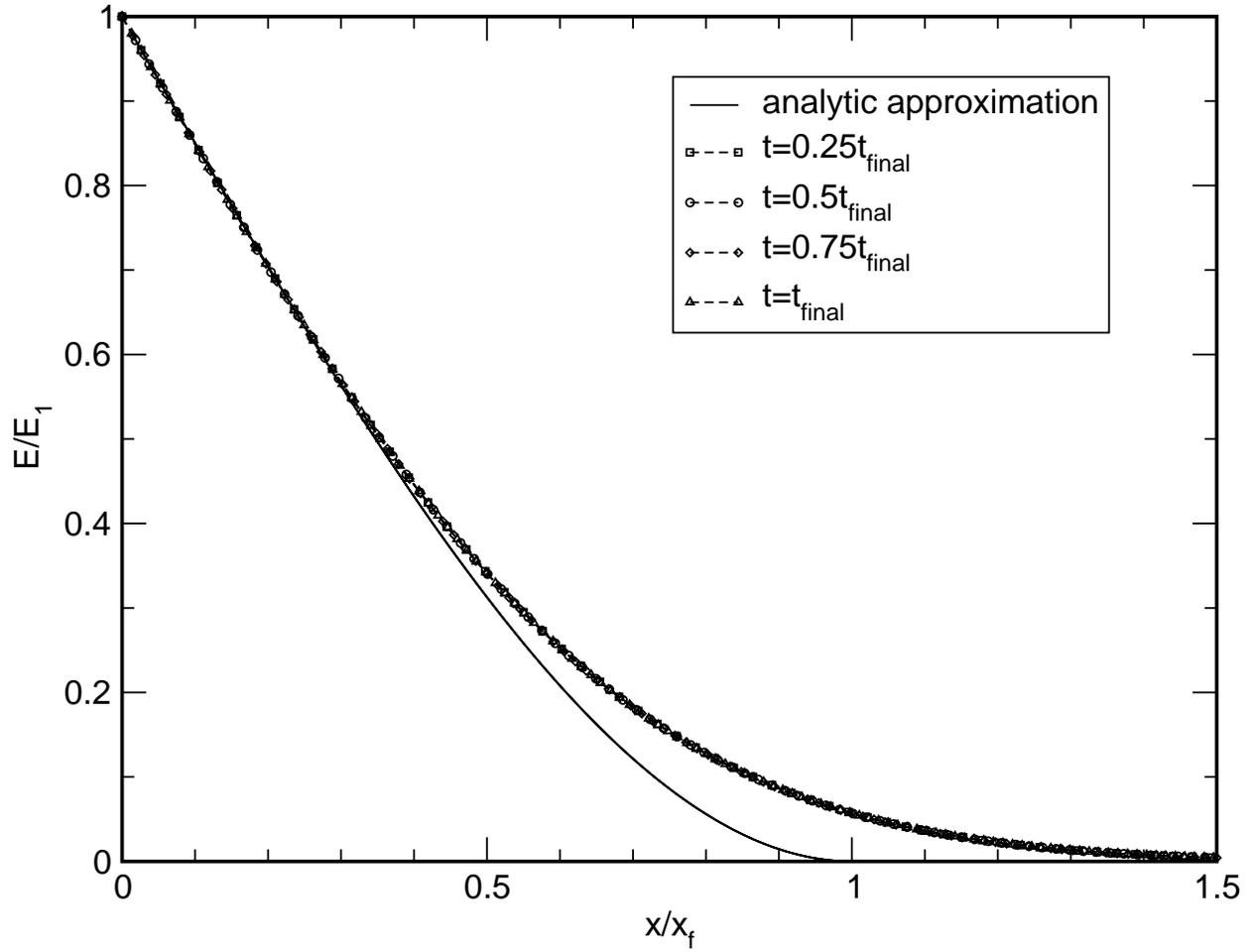}
\caption{Profile of the radiation front at 4 different times for
a Marshak wave propagating through a constant opacity medium.
Also shown is an analytic approximation of the profile.
\label{fig:marsh_pro}}
\end{figure}

\begin{figure}
\plotone{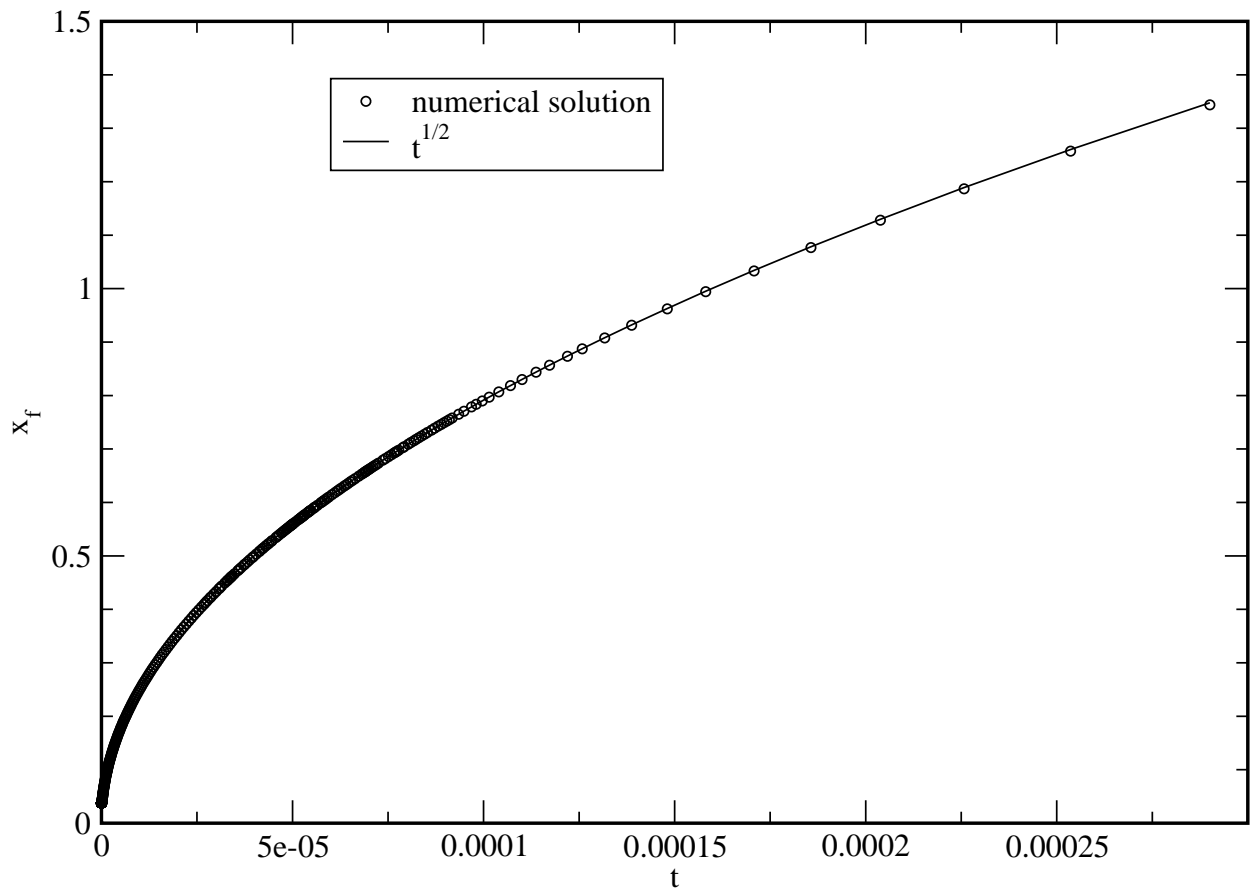}
\caption{Plot of the radiation front position as a function of time for a 
Marshak wave propagating through a constant opacity medium.  The data
is fit with a $t^{1/2}$ curve, which is the expected relation.
\label{fig:marsh_time}}
\end{figure}

\begin{figure}
\plotone{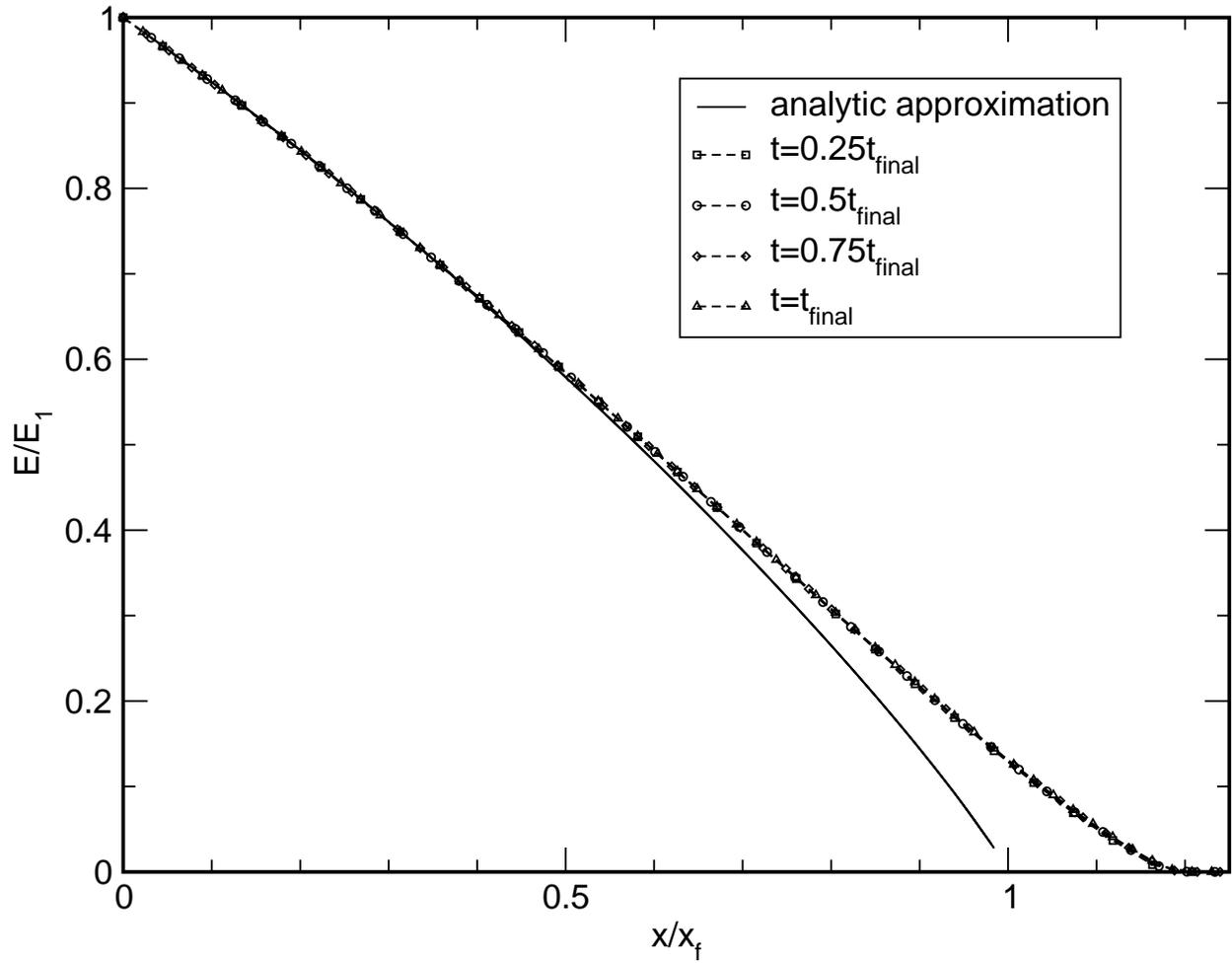}
\caption{Same as Figure \ref{fig:marsh_pro} except for a temperature-dependent
opacity of the form $\sigma_r \propto T^{-3}$.
\label{fig:marsh_pro_T}}
\end{figure}

\begin{figure}
\plotone{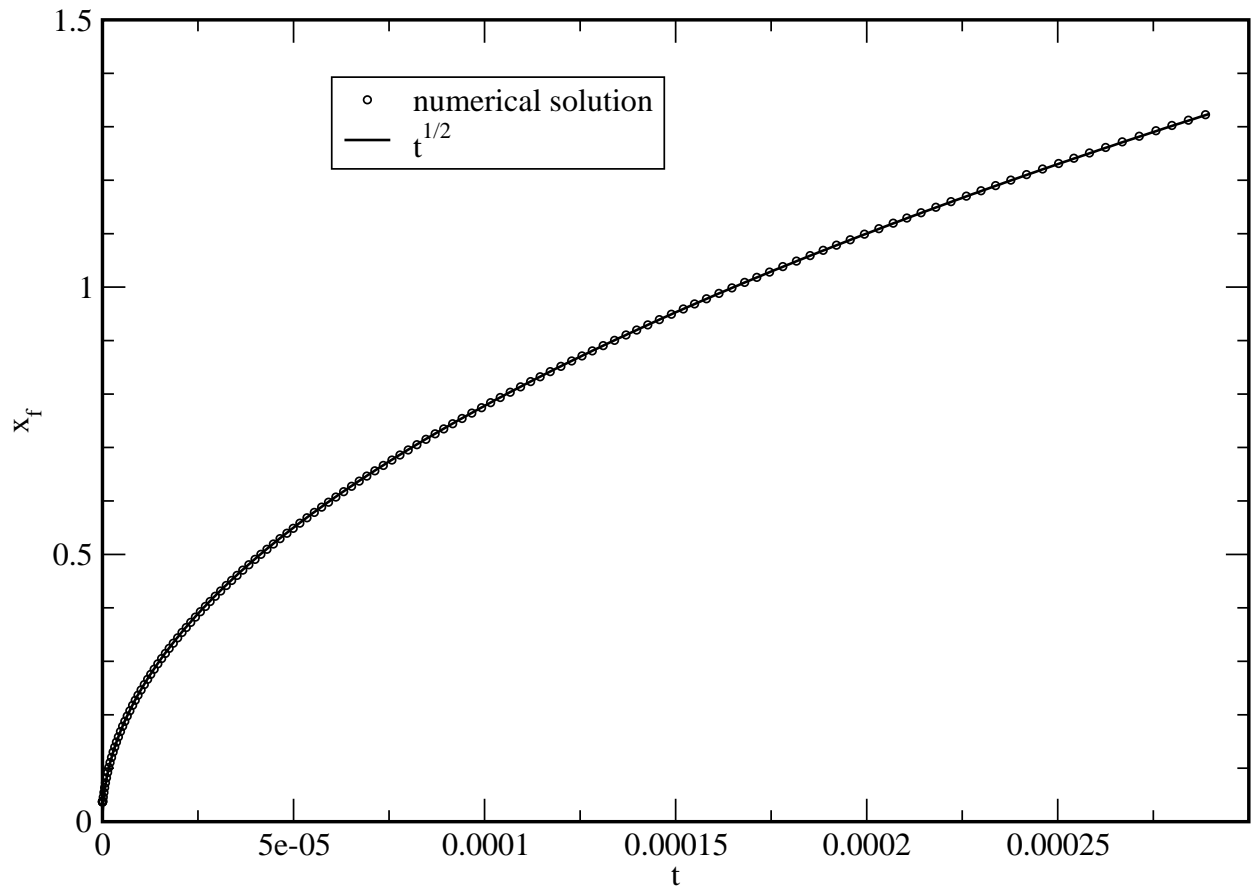}
\caption{Same as Figure \ref{fig:marsh_time} except for a 
temperature-dependent opacity.
\label{fig:marsh_time_T}}
\end{figure}

In this work we consider two cases: a constant opacity with $\sigma_r=\kappa_0$,
and a temperature-dependent opacity with $\sigma_r=\kappa_0(T/T_0)^{-m}$ 
and $m=3$.  The computational grid is set to approximately 50 times the
mean-free path ($l=1/3\rho \sigma_r$), and an appropriate stopping time
$t_{final}$ is defined using equation (\ref{eqn:Marsh_front}).
The problem is initialized with $\kappa_0=1$ cm$^2$ g$^{-1}$,
$\rho=1$ g cm$^{-3}$, $T_0=10^4$ K, $T_1=10^6$ K,  
and the grid is discretized with 400 zones.
Figure \ref{fig:marsh_pro} shows the profile of 
the radiation front for the constant opacity case 
normalized to the analytic radiation front position
with $\eta=1.1$.  
The numerical results, plotted at (0.25, 0.5, 0.75, 1)$\times t_{final}$, 
nicely display the self-similar nature of this 
solution.
Figure \ref{fig:marsh_time} plots the numerical radiation front 
location as a function of time.  Here we define the numerical front to be
located at the half-maximum, although our results do not depend strongly
upon this choice.  The data is well fit with a $t^{1/2}$ curve.  
Figures \ref{fig:marsh_pro_T} and 
\ref{fig:marsh_time_T} present similar results for the temperature-dependent 
opacity.  The analytic curve in Figure \ref{fig:marsh_pro_T} is computed with
$\eta=1.04$.

\subsection{$\Gamma=2$ Polytrope}
\label{sec:polytrope}

Here we test the linear matrix methods used in solving Poisson's
equation for the gravitational potential of a compact
self-gravitating source, and the ability of the code to maintain
a balance between gravitational and pressure
support forces in three dimensions. 
For this test we adopt an adiabatic polytrope star with $\Gamma=2$
in spherical symmetry and hydrostatic equilibrium with
solutions for the gas pressure, density and gravitational potential
\begin{eqnarray}
P    &=& 2\pi G \alpha^2 \rho^2 , \label{eqn:nstarp} \\
\rho &=& \frac{\alpha \rho_c}{r} \sin{\left(\frac{r}{\alpha}\right)} , \\
\phi &=& -\frac{G M}{R_s} - \frac{4\pi G \rho_c \alpha^3}{r} 
                            \sin{\left(\frac{r}{\alpha}\right)}  \label{eqn:nstarphi} ,
\end{eqnarray}
where $\alpha = (M/(4\pi^2 \rho_c))^{1/3}$
for radii $r \le R_s$,  and $R_s = \alpha \pi$ is the outer surface radius.
Also, $\phi = -GM/r$ with negligible density
and pressure ($< 10^{-4}$ of the maximum central values)
outside the star at radii $r>R_s$.
These tests are carried out
for characteristic neutron star parameters with total mass
$M = 1.4 M_{\odot}$, central density $\rho_c=2.5\times10^{15}$ g cm$^{-3}$,
and radius 5.6 km.

\begin{figure}
\plotone{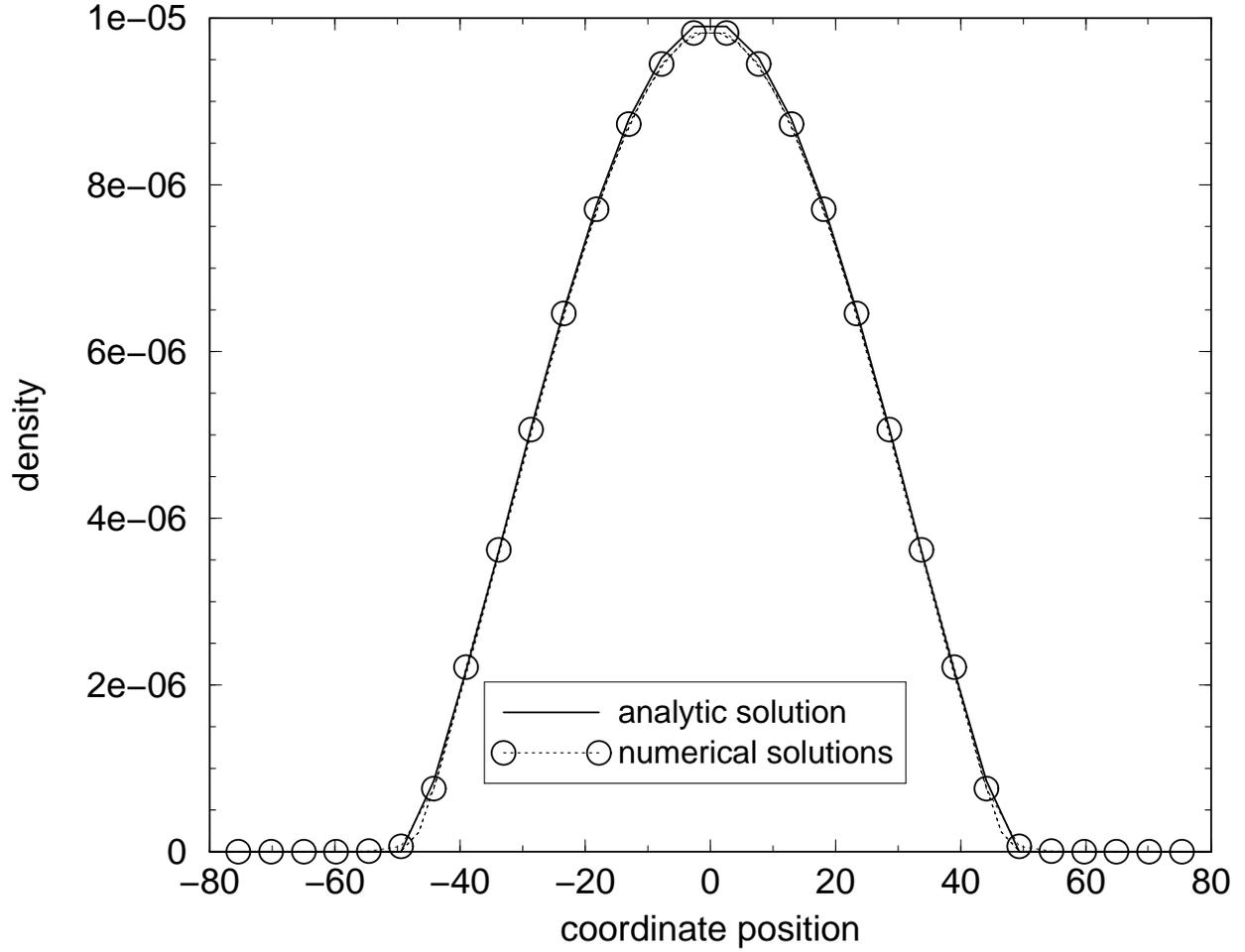}
\caption{Density profiles for a $48^3$ zone test of the 
spherical hydrostatic
$\Gamma=2$ polytrope solution. Profiles along the $x$, $y$, $z$
and diagonal directions through the origin are displayed at the
initial time (analytic solution represented by the solid line), 
and after two sound crossing times (open circles).
\label{fig:nstar}}
\end{figure}

We ran a sequence of three simulations at different grid resolutions
over two sound crossing times,
$t=2T_s=2 R_s/\sqrt{2 \pi G \alpha^2 \rho_c}$,
with monopole boundary conditions to find mean relative
errors of 0.0687, 0.0209, and 0.0101 
for resolutions $12^3$, $24^3$, and $48^3$ cells respectively.
Monopole boundary conditions are implemented by computing
the total mass $M_T$ and mass centroid coordinates 
$x^k_c = \sum_{cells} \rho x^k \Delta x\Delta y \Delta z/M_T$ 
in the computational domain. The potential field is then set to
$\phi = - G M_T/\sqrt{\sum_k (x^k_o-x^k_c)^2}$ 
at the outer boundary $x^k = x^k_o$.
Figure \ref{fig:nstar} shows spatial profiles
in density for the $48^3$ case along the $x$-axis
at the initial time (the analytic solution),
and along four separate directions at $t=2T_s$: the $x$,
$y$, $z$ and diagonal lines running through the origin.
The data is displayed in dimensionless code units with
length scale $\widetilde{L} = R_s/N$ where $N=48$
is the number of interior cells along an axis, and density scale
$\widetilde{D} = M_{\odot}/\widetilde{L}^3$.
We observe no significant break in symmetry in the solutions,
and the central peak density  is maintained to good
accuracy.

\subsection{$\Gamma=2$ Polytrope with Radiation Pressure}
\label{sec:polytroperad}

The hydrostatic polytrope solution in \S\ref{sec:polytrope} can
be generalized to include a radiation field and radiation pressure,
thus providing a useful test of the coupling between
gravity, fluid pressure, and radiation pressure. It also exercises
the multiphysics operator splitting scheme in testing
the ability of the code to maintain a balance between the
three different self-consistently generated forces in three dimensions.

The solution (\ref{eqn:nstarp}) - (\ref{eqn:nstarphi}) is easily extended
to include the effect of radiation pressure by assuming $P_R=\beta P$,
where $\beta$ is a constant. This simplifies the solution considerably
and allows for an effective (hydrodynamic plus radiation) pressure to
counteract gravity with the same radial dependence as in the
pure hydrodynamics case. The radiation energy equation (\ref{eqn:enrad})
reduces, in the spherical, static, and thermal equilibrium limits,
to $r^{-2}\partial_r (r^2 D \partial_r E) = 0$, which can be solved
trivially if the diffusion coefficient is set to $D=k_0/(r^2\partial_r E)$,
where $k_0$ is a constant. The complete solution including radiation
pressure and radially dependent opacity is
\begin{eqnarray}
P    &=& \frac{2\pi G \alpha^2}{1+\beta} \rho^2 , \\
\rho &=& \frac{\alpha \rho_c}{r} \sin{\left(\frac{r}{\alpha}\right)} , \\
\phi &=& -\frac{G M}{R_s} - \frac{4\pi G \rho_c \alpha^3}{r} 
                            \sin{\left(\frac{r}{\alpha}\right)}  , \\
P_R  &=& \frac{E}{3} = \frac{2\pi G \beta \alpha^2}{1+\beta} \rho^2 , \\
\sigma_r &=& \frac{4\pi G c \beta \rho_c \alpha^3}{k_0 (1+\beta)}
\left(\frac{r}{\alpha} \cos\left(\frac{r}{\alpha}\right) 
      - \sin\left(\frac{r}{\alpha}\right) \right) ,
\end{eqnarray}
where $\sigma_r = c/(3\rho D)$ is the Rosseland mean opacity,
and the density and gravitational potential are unchanged.

\begin{figure}
\plotone{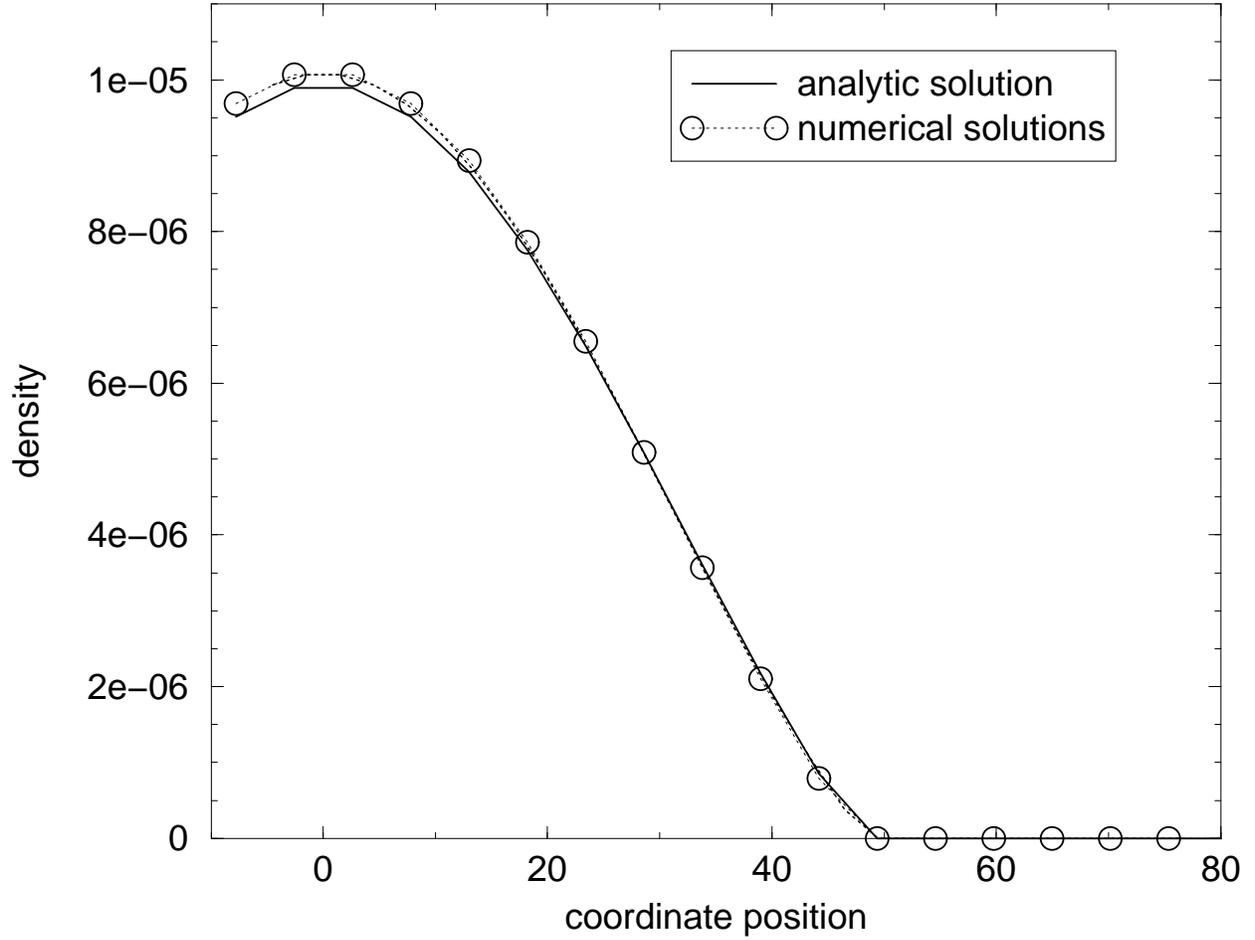}
\caption{Density profiles for a $24^3$ zone test of the
spherical hydrostatic
$\Gamma=2$ polytrope solution with radiation pressure. 
Profiles along the $x$, $y$, $z$
and diagonal directions through the origin are displayed at the
initial time (analytic solution represented by the solid line),
and after 0.5 sound crossing times (open circles).
\label{fig:nstarrad}}
\end{figure}

We ran a sequence of three simulations at different grid resolutions
over half a sound crossing time and in an octant to properly
specify the constant radiation boundary conditions across the star profile.
The parameters are the same as the tests in \S\ref{sec:polytrope},
but with the addition of $k_0=1$ and $\beta=1$ to give equal significance
to the radiation and hydrodynamic pressures.
We find mean relative errors of 0.148, 0.0405, and 0.0138
for resolutions $6^3$, $12^3$, and $24^3$ cells respectively,
demonstrating roughly second order convergence.
Figure \ref{fig:nstarrad} shows spatial profiles
in density for the $24^3$ case at the initial time (the analytic solution),
and along the $x$, $y$, $z$ and diagonal directions running through the origin
at $t=T_s/2$.

\subsection{Astrophysical Jets}
\label{sec:jets}

In this last section we perform simulations of astrophysical jets as a final test
of our code. A beam of low density material ($n_J = 10^{-3}~\mbox{cm}^{-3}$)
with radius $R_J$ is injected into a homogeneous higher density
($n_A = 10 n_J$) ambient medium along the $+z$ axis.
Within a cylindrical radius $\sqrt{x^2+y^2} \le R_J = 500$ parsecs
in the $z=0$ plane, a constant inflow velocity of $V_J=10^9$ cm s$^{-1}$
is maintained as a boundary condition. Simple outflow
conditions are imposed at all other external boundaries.
The jet material and ambient medium are initialized in pressure equilibrium, 
with an ambient temperature of $10^7~K$. An ideal gas equation of state
with adiabatic index $\Gamma=5/3$ is used for both materials. The Mach number
for these parameters is $M=V_J/c_s \approx 36$, 
where $c_s$ is the sound speed of the
background material into which the beam flows.

\begin{figure}
\epsscale{0.5}
\plotone{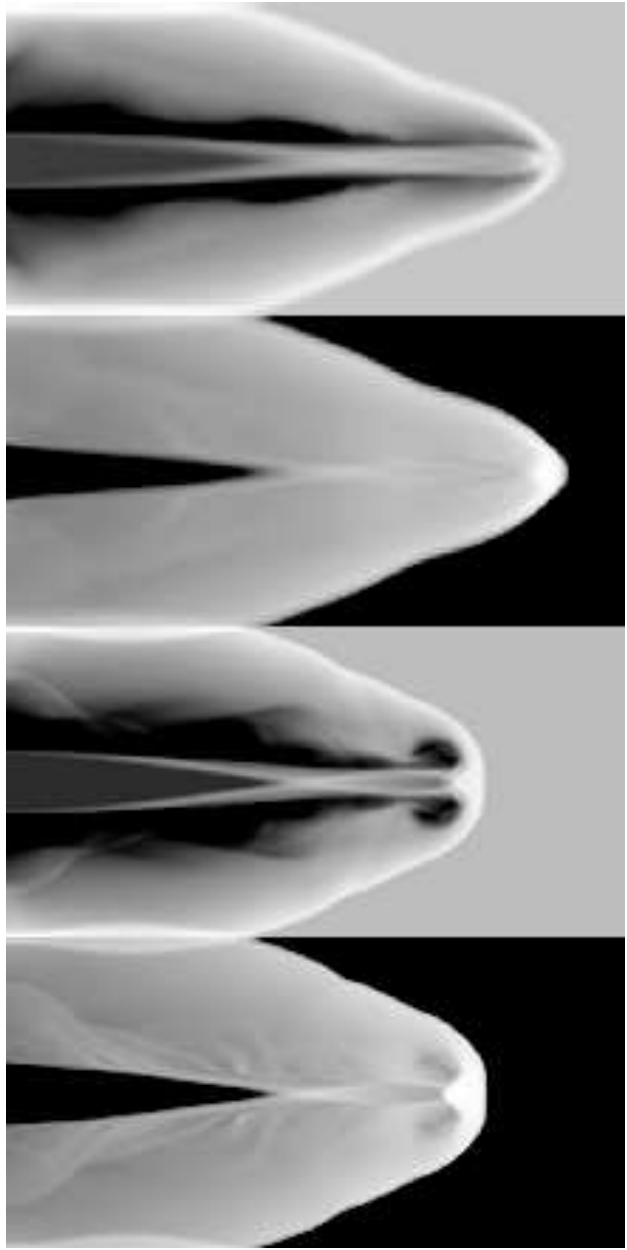}
\caption{Cross sections in the $y=0$ plane of two 
simulations of jets propagating to the right along
the positive $z$ axis. From top to bottom: low resolution
density, low resolution internal energy, high
resolution density, high resolution internal energy.
Six (twelve) cells are used to resolve the jet
radius in the low (high) resolutions.
\label{fig:jets}}
\end{figure}

Figure \ref{fig:jets} shows results at time $t=54~R_J/V_J$ 
from two calculations: one using a low resolution grid 
($64\times64\times128$ cells)
with 6 zones to cover the jet radius, and a high resolution grid
($128\times128\times256$ cells) with 12 zones/$R_J$. 
The physical box dimensions
in both cases are set to $10R_J\times10R_J\times20R_J$. The results,
particularly the high resolution case, clearly show all the morphological 
elements of astrophysical jets \citep{Norman82}: 
a supersonic beam that ends in a bow shock,
a cocoon composed of shock heated jet material, a working surface separating
jet and shocked ambient gas, internal shock interactions, and the growth
of Kelvin-Helmholtz instabilities.

Equating the ram pressure from the jet front and the equivalent pressure
from the ambient medium yields
\begin{equation}
\overline{V}_s = \frac{V_J}{1+\sqrt{\rho_A/\rho_J}} ,
\label{eqn:jetvel}
\end{equation}
for an estimate of the velocity of the bow shock through the ambient medium,
neglecting multi-dimensional effects. The sensitivity of the leading shock
velocity $V_s$ to grid resolution is clear from Figure \ref{fig:jets}.
In particular, we find, by tracking the bow shock position, velocities
of $V_s = 0.32 V_J$ and $0.26 V_J$ for the low and high resolution
runs respectively. In comparison, equation (\ref{eqn:jetvel}) predicts
$\overline{V}_s = 0.24 V_J$. Higher resolution allows more accurate modeling of
the bow shock and working surface which effectively
broadens the jet, and resolves to a greater extent 3D instabilities
and internal shock interactions, all of which contribute to slowing the shock.
These results are generally consistent with those of \citep{Massaglia96}, who
find jets with similar hypersonic Mach numbers
and density ratios propagate
at near (and greater than) unit efficiencies, defined by
$V_s/\overline{V}_s \gtrsim 1$.

\section{Summary}
\label{sec:summary}

We have developed a new multidimensional, multiphysics code (Cosmos)
that can be applied to a broad range of astrophysical problems,
from highly relativistic scalar-field dominated applications in
early universe cosmology, to black-hole accretion and multiphase
Newtonian galactic dynamics.  In this contribution, we presented the
radiation-chemo-hydrodynamics
equations solved in the Newtonian limit, along with numerous
tests to gauge the accuracy and stability of the code in various multiphysics
modes. In a companion paper \citep{FMAL02}, Cosmos is applied to the problem of
supernova-enrichment in dwarf-spheroidal galaxies, which utilizes
many of the capabilities (robust shock capturing, radiative cooling
flows, multiphase fluids, hydrodynamic instabilities,
and gravitational potentials) presented and tested here.
This work also complements an earlier paper \citep{Anninos03} 
in which we presented
the general relativistic equations solved in Cosmos. There we provided
more detailed descriptions of our numerical methods
for the different energy and algorithmic formulations,
together with numerical tests for highly relativistic 
hydrodynamical systems and black hole accretion.

\begin{acknowledgements}
This work was performed
under the auspices of the U.S. Department of Energy by
University of California, Lawrence
Livermore National Laboratory under Contract W-7405-Eng-48.
\end{acknowledgements}

\clearpage

\clearpage

\begin{table}[tbm]
\begin{tabular}{ll}
\hline\hline
Primordial Chain & Molecular Chain \\
\hline
k1:\quad\ (1)\hskip10pt  $\mathrm{H}       +  e    
              \rightarrow  \mathrm{H}^+     +  2e$               &
k7:\quad\ (7)\hskip10pt  $\mathrm{H}       +  e    
              \rightarrow  \mathrm{H}^-     +  \gamma$           \\
k2:\quad\ (2)\hskip10pt  $\mathrm{H}^+     +  e    
              \rightarrow  \mathrm{H}       +  \gamma$           &
k8:\quad\ (8)\hskip10pt  $\mathrm{H}       +  \mathrm{H}^-  
              \rightarrow  \mathrm{H}_2     +  e$                \\
k3:\quad\ (3)\hskip10pt  $\mathrm{He}      +  e    
              \rightarrow  \mathrm{He}^{+}  +  2e$               &
k9:\quad\ (9)\hskip10pt  $\mathrm{H}       +  \mathrm{H}^+  
              \rightarrow  \mathrm{H}_2^+   +  \gamma$           \\
k4:\quad\ (4)\hskip10pt  $\mathrm{He}^+    +  e    
              \rightarrow  \mathrm{He}      +  \gamma$           &
k10:\quad (10)\hskip10pt  $\mathrm{H}_2^+   +  \mathrm{H}    
              \rightarrow  \mathrm{H}_2     +  \mathrm{H}^+ $    \\
k5:\quad\ (5)\hskip10pt  $\mathrm{He}^+    +  e    
              \rightarrow  \mathrm{He}^{++} +  2e$               &
k11:\quad (11)\hskip10pt  $\mathrm{H}_2     +  \mathrm{H}^+  
              \rightarrow  \mathrm{H}_2^+   +  \mathrm{H} $      \\
k6:\quad\ (6)\hskip10pt  $\mathrm{He}^{++} +  e    
              \rightarrow  \mathrm{He}^+    +  \gamma$           &
k12:\quad (12)\hskip10pt  $\mathrm{H}_2     +  e    
              \rightarrow  2\mathrm{H}      +  e$                \\
\ \ &
k13:\quad (13)\hskip10pt  $\mathrm{H}_2     +  \mathrm{H}    
               \rightarrow  3\mathrm{H}$                          \\
\ \ &
k14:\quad (14)\hskip10pt  $\mathrm{H}^-     +  e    
               \rightarrow  \mathrm{H}       +  2e$               \\
\ \ &
k15:\quad (15)\hskip10pt  $\mathrm{H}^-     +  \mathrm{H}    
               \rightarrow  2\mathrm{H}      +  e$                \\
\ \ &
k16:\quad (16)\hskip10pt  $\mathrm{H}^-     +  \mathrm{H}^+  
               \rightarrow  2\mathrm{H}$                          \\
\ \ &
k17:\quad (17)\hskip10pt  $\mathrm{H}^-     +  \mathrm{H}^+  
               \rightarrow  \mathrm{H}_2^+   +  e$                \\
\ \ &
k18:\quad (18)\hskip10pt  $\mathrm{H}_2^{+} +  e   
               \rightarrow  2\mathrm{H}$                          \\
\ \ &
k19:\quad (19)\hskip10pt  $\mathrm{H}_2^{+} +  \mathrm{H}^- 
               \rightarrow  \mathrm{H}_2     +  \mathrm{H}$       \\
\hline
k20:\quad (20)\hskip10pt  $\mathrm{H}       +  \gamma    
               \rightarrow  \mathrm{H}^+     +  e   $             &
k23:\quad (23)\hskip10pt $\mathrm{H}^-      +  \gamma    
               \rightarrow \mathrm{H}        +  e$                \\
k21:\quad (21)\hskip10pt  $\mathrm{He}      +  \gamma    
               \rightarrow  \mathrm{He}^+    +  e   $             &
k24:\quad (24)\hskip10pt $\mathrm{H}_2      +  \gamma    
               \rightarrow \mathrm{H}_2^{+}  +  e$                \\
k22:\quad (22)\hskip10pt  $\mathrm{He}^+    +  \gamma    
               \rightarrow  \mathrm{He}^{++} +  e   $             &
k25:\quad (25)\hskip10pt $\mathrm{H}_2^+    +  \gamma    
               \rightarrow \mathrm{H}        +  \mathrm{H}^+ $    \\
\ \ &
k26:\quad (26)\hskip10pt $\mathrm{H}_2^+    +  \gamma    
               \rightarrow 2\mathrm{H}^+     +  e$                \\
\ \ &
k27:\quad (27)\hskip10pt $\mathrm{H}_2      +  \gamma    
               \rightarrow 2\mathrm{H}$                           \\
\hline\hline
\end{tabular}
\caption{
Chemical gas phase reactions modeled in Cosmos, grouped by
primordial versus molecular chains, and collisional
versus photoreactive processes.
The corresponding rate coefficients are referred to as $k_i$
in the main text, where $i$ is the reaction number defined in this table.
For a more detailed description of the chemistry and for
explicit formulas used in defining the kinetic and cooling coefficients, see
\protect{\cite{Abel97}} and \protect{\cite{Anninos97}}.
}
\label{tab:chemistry}
\end{table}

\begin{deluxetable}{cccc}
\tablewidth{0pt}
\tablecaption{L-1 Norm Errors in density, pressure, and velocity for the
Sod shock tube test \label{tab:Sod}.}
\tablehead{
\colhead{Grid} &
\colhead{$\Vert E(\rho) \Vert_1$} &
\colhead{$\Vert E(P) \Vert_1$} &
\colhead{$\Vert E(V) \Vert_1$}
}
\startdata
32 & $1.62 \times 10^{-2}$ & $1.67 \times 10^{-2}$ & $4.82 \times 10^{-2}$ \\
64 & $8.99 \times 10^{-3}$ & $8.68 \times 10^{-3}$ & $2.37 \times 10^{-2}$ \\
128 & $4.91 \times 10^{-3}$ & $4.51 \times 10^{-3}$ & $1.34 \times 10^{-2}$ \\
$64^3$ & $1.02 \times 10^{-2}$ & $1.04 \times 10^{-2}$ & $6.92 \times 10^{-3}$
\enddata
\end{deluxetable}

\end{document}